\documentclass[a4paper,oneside,10pt]{amsart}

\usepackage[T1]{fontenc}
\usepackage{amsmath,amssymb,epsf}
\usepackage{amsfonts,amsbsy,amscd,stmaryrd}
\usepackage{latexsym,amsopn}
\usepackage{amsthm}
\usepackage{esint}
\usepackage{mathrsfs}
\usepackage{float}

\usepackage{graphicx}
\usepackage{color}
\definecolor{Blue}{rgb}{0.,0.,1.}
\definecolor{Red}{rgb}{1.,0.,0.}

\newcounter{smallarabics}
\newenvironment{arabicenumerate}
{\begin{list}{{\normalfont\textrm{(\arabic{smallarabics})}}}
  {\usecounter{smallarabics}\setlength{\itemindent}{0cm}
   \setlength{\leftmargin}{5ex}\setlength{\labelwidth}{4ex}
   \setlength{\topsep}{0.75\parsep}\setlength{\partopsep}{0ex}
   \setlength{\itemsep}{0ex}}}
{\end{list}}

\newcounter{smallroman}

\let\origmaketitle\maketitle
\def\maketitle{
  \begingroup
  \def\uppercasenonmath##1{} 
  \let\MakeUppercase\relax 
	\origmaketitle
  \endgroup
}

\newcommand{\ben}{\begin{arabicenumerate}}  
\newcommand{\een}{\end{arabicenumerate}}

\def\init{\setcounter{equation}{0}}

\newtheorem{theorem}{Theorem}[section]

\newtheorem{proposition}[theorem]{Proposition}
\newtheorem{lemma}[theorem]{Lemma}
\newtheorem{definition}[theorem]{Definition}

\newtheorem{remark}[theorem]{Remark}
\newtheorem{example}[theorem]{Example}
\newcommand{\beq}{\begin{equation}}
\newcommand{\eeq}{\end{equation}}

\newcommand{\bex}{\begin{example}}
\newcommand{\eex}{\end{example}}
\def\bel{\begin{lemma}}
\def\eel{\end{lemma}}
\def\bet{\begin{theorem}}
\def\eet{\end{theorem}}
\def\bed{\begin{definition}}
\def\eed{\end{definition}}
\def\ber{\begin{remark}}
\def\eer{\end{remark}}

\def\rr{{\mathbb R}}

\def\cc{{\mathbb C}}

\def\bar{\overline}

\def\cinf{C^\infty}
\def\c0inf{C_0^\infty}

\def\proof{
\noindent{\bf Proof.}\ \ }

\def\sh{{\rm sh}}

\usepackage{calrsfs}
\DeclareMathAlphabet{\pazocal}{OMS}{zplm}{m}{n}
\DeclareMathAlphabet{\mathsfsl}{OMS}{cmss}{m}{n}

\DeclareSymbolFont{altletters}  {OML}{zplm}{m}{it}
\DeclareMathSymbol{\altdelta}{\mathalpha}{altletters}{"0E}
\DeclareMathSymbol{\alteta}{\mathalpha}{altletters}{"11}

\def\cY{{\pazocal Y}}

\def\cS{{\pazocal S}}

\def\cD{{\pazocal D}}

\def\cN{{\pazocal N}}

\def\CCR{{\rm CCR}}

\def\a{{\rm a}}

\def\i{{\rm i}}

\DeclareMathOperator{\Dom}{Dom}

\def\qed{$\Box$\medskip}

\DeclareMathOperator{\Ker}{Ker}

\def\t{\tilde{}}

\def \p{ \partial}
\def\12{\frac{1}{2}}
\def\14{\frac{1}{4}}

\def\supp{{\rm supp}}

\def\e{{\rm e}}

\newcommand{\one}{\boldsymbol{1}}

\def\coinf{C_{\rm c}^\infty}

\def\c{{\pazocal }}

\def\cX{{\pazocal X}}

\def\t{\tilde{}}

\def\12{\frac{1}{2}}

\def\supp{{\rm supp}}

\def\e{{\rm e}}

\def\bep{\begin{proposition}}
\def\eep{\end{proposition}}

\def\b{{\rm b}}

\newcommand{\mat}[4]{\begin{pmatrix}#1 &#2  \\ #3 &#4 \end{pmatrix}}
\newcommand{\col}[2]{\begin{pmatrix}#1 \\#2\end{pmatrix}}

\def\CARal{{\rm C\hskip 0.25 em \hbox{\raise 1.72 ex 
\hbox{$\scriptscriptstyle\rm al$}\kern -0.57 em A}R}}

\def\t{{\scriptscriptstyle\#}}
\def\otimesal{\mathop{\hbox{\raise 1.5 ex
  \hbox{$\scriptscriptstyle\rm al$}
\kern -0.92 em \hbox{$\otimes$}}}}
\def\oplusal{\mathop{\hbox{\raise 1.5 ex
  \hbox{$\scriptscriptstyle\rm al$}
\kern -0.92 em \hbox{$\oplus$}}}}
\def\Gammal{\hbox{\raise 1.68 ex 
\hbox{$\scriptscriptstyle\rm al$}\kern -0.50 em $\Gamma$}}
\def\Bal{\hbox{\raise 1.68 ex 
\hbox{$\scriptscriptstyle\rm  al$}\kern -0.50 em $B$}}
\def\CARal{{\rm C\hskip 0.25 em \hbox{\raise 1.72 ex 
\hbox{$\scriptscriptstyle\rm al$}\kern -0.57 em A}R}}
\def\t{{\scriptscriptstyle\#}}

\DeclareMathAlphabet{\mathpzc}{OT1}{pzc}{m}{it}

\DeclareSymbolFont{boldoperators}{OT1}{cmr}{bx}{n}
\SetSymbolFont{boldoperators}{bold}{OT1}{cmr}{bx}{n}

\makeatletter
\newcommand*{\defeq}{\mathrel{\rlap{%
                     \raisebox{0.3ex}{$\m@th\cdot$}}%
                     \raisebox{-0.3ex}{$\m@th\cdot$}}%
                     =}
                     
\makeatother
\makeatletter
\newcommand*{\eqdef}{=\mathrel{\rlap{%
                     \raisebox{0.3ex}{$\m@th\cdot$}}%
                     \raisebox{-0.3ex}{$\m@th\cdot$}}%
                     }
\makeatother

\DeclareMathAlphabet{\mathpzc}{OT1}{pzc}{m}{it}

\def\WF{{\rm WF}}

\newcommand{\bea}{\begin{aligned}}
\newcommand{\beal}{\begin{array}{l}}
\newcommand{\eeal}{\end{array}}
\newcommand{\eea}{\end{aligned}}

\def\pe{\overline{\p}}

 \def\cotn{\coth(\frac{\beta}{2}\epsilon)}
 \def\shn{\sh^{-1}(\frac{\beta}{2}\epsilon)}
 \def\cotp{\coth(\frac{\beta}{2}\epsilon_{+})}
 \def\shp{\sh^{-1}(\frac{\beta}{2}\epsilon_{+})}
 \def\epn{\epsilon}
 \def\epp{\epsilon_{+}}
\def\dito{\!\cdot\!}
\def\calderon{Calder\'{o}n }
\def\cB{\mathcal{B}}

\def\pe{\overline{\p}}
\def\bS{\mathbb{S}}
\newcommand{\tra}[1]{\mskip-6mu\upharpoonright_{#1}\mskip+4mu}

\begin{document}
\title[Hartle-Hawking-Israel states and \calderon projectors]{\large On the Hartle-Hawking-Israel states 
 \\ for spacetimes with static bifurcate Killing horizons}
\author{}
\address{Universit\'e Paris-Sud XI, D\'epartement de Math\'ematiques, 91405 Orsay Cedex, France}
\email{christian.gerard@math.u-psud.fr}
\date{August 2016}
\author{\normalsize Christian \textsc{G\'erard} }
\begin{abstract}We revisit the construction by Sanders \cite{S} of the Hartle-Hawking-Israel state for a free quantum Klein-Gordon field on a spacetime with a static, bifurcate Killing horizon and a wedge reflection. Using the notion of the \calderon projector for elliptic boundary value problems and pseudodifferential calculus on manifolds, we  give a short proof of its Hadamard property.
\end{abstract}

\maketitle

\section{Introduction}\init
Let $(M, g)$ be a globally hyperbolic spacetime, with a {\em bifurcate Killing horizon}, see \cite{KW}, \cite{S} or Subsect. \ref{sec1.1} for precise definition.  The bifurcate Killing horizon $\mathcal{H}$ is generated by the {\em bifurcation surface} $\mathcal{B}= \{x\in M : V(x)=0\}$, where $V$ is the Killing vector field. It allows to split $(M,g)$ into four globally hyperbolic regions, the {\em right/left wedges} ${\mathcal M}^{+}$, ${\mathcal M}^{-}$ and the {\em future/past cones} $\mathcal{F}$, $\mathcal{P}$, each invariant under the flow of  $V$.  An important object related with the Killing horizon  $\mathcal{H}$ is its {\em surface gravity} $\kappa$, which is a scalar, constant over all of $\mathcal{H}$.

Let us consider on $(M,g)$ a free quantum Klein-Gordon field associated to the Klein-Gordon equation
\[
-\Box_{g}\phi(x)+ m(x)\phi(x)=0,
\]
where $m\in\cinf(M, \rr)$, $m(x)>0$ is invariant under $V$, and its associated free field algebra.

If  $V$ is {\em time-like} in $({\mathcal M}^{+}, g)$,  ie if $({\mathcal M}^{+}, g, V)$ is a stationary spacetime, there exists (see \cite{S2}) for any $\beta>0$ a {\em thermal state} $\omega_{\beta}^{+}$  at temperature $\beta^{-1}$ with respect to the group of Killing isometries of $({\mathcal M}^{+}, g)$ generated by $V$. 

It was conjectured by Hartle and Hawking \cite{HH} and Israel \cite{I} that  if $\beta= \dfrac{2\pi}{\kappa}$  is the {\em inverse Hawing temperature}, denoted by $\beta_{\rm H}$ in the sequel,  then  $\omega_{\beta}^{+}$ can be extended to the whole of $M$ as a pure state, invariant under $V$, the {\em Hartle-Hawking-Israel state}, denoted in the sequel by $\omega_{\rm HHI}$.

The rigorous construction of  the HHI state was first addressed by Kay in \cite{K4}, who constructed the HHI state  in the Schwarzschild  double wedge of the Kruskal spacetime.  In such a double wedge, the HHI state is a {\em double KMS state}, see \cite{K2, K3}.  Later Kay and Wald  \cite{KW} considered the more general case of spacetimes with a bifurcate  Killing horizon, and study general properties  of stationary states on this class of spacetimes. They   emphasized in particular the importance of the {\em Hadamard condition}.  They proved that a specific sub-algebra of the free field algebra has at most one   state invariant under $V$ and Hadamard. They also showed that  if $M$ admits a {\em wedge reflection} (see Subsect. \ref{sec1.2}) the restriction of such a state to ${\mathcal M}^{+}$ will necessarily be a $\beta_{\rm H}-$KMS state. These results were later improved in \cite{K}.

The existence of such a state, ie of the HHI state, was however not proved in \cite{HH}. The first proof of the existence of $\omega_{\rm HHI}$ was given by Sanders in the remarkable paper \cite{S}, if the bifurcate Killing horizon is static, ie if $V$ is static in ${\mathcal M}^{+}$, assuming also the existence of a wedge reflection.  Sanders showed that there exists a unique Hadamard state $\omega_{\rm HHI}$ on $M$ extending the {\em double} $\beta_{\rm H}-${\em KMS state} $\omega_{\beta}$ on ${\mathcal M}^{+}\cup {\mathcal M}^{-}$. The double $\beta_{H}-$KMS state $\omega_{\beta}$ is a pure state on ${\mathcal M}^{+}\cup {\mathcal M}^{-}$ which is the natural extension of $\omega_{\beta}^{+}$ defined using the wedge reflection, see \cite{K2, K3}. It is an exact geometrical analog of  the Fock vacuum vector  in the Araki-Woods representation of a thermal state.

\subsection{Content of the paper}
In this paper we revisit the construction  in \cite{S} of the Hartle-Hawking-Israel state in a spacetime with a static bifurcate Killing horizon.  Using the notion of the {\em Calder\'{o}n projector} (see Sect. \ref{sec3}), which is a standard tool in elliptic boundary value problems, we significantly shorten the proof of the Hadamard property of $\omega_{\rm HHI}$. 

In \cite{S} the fact that $\omega_{\rm HHI}$ is Hadamard was proved by a careful comparison of the Hadamard parametrix construction for the D'Alembertian $- \Box_{g}+ m$ associated to the Lorentzian metric $g$ and  for the Laplacian $- \Delta_{\hat{g}}+ m$ associated to the Riemannian metric $\hat{g}$ obtained from $g$ by  Wick rotation in the Killing time coordinate. 

In our paper we avoid working with the spacetime covariances of states and instead  systematically work with the {\em Cauchy surface covariances} (see Subsect. \ref{csc}) associated with a Cauchy surface $\Sigma$ containing the bifurcation surface $\mathcal{B}$. 

It turns out that the Cauchy surface covariances $\lambda^{\pm}$ of  the double $\beta-$KMS state 
$\omega_{\beta}$ are related to a  Calder\'{o}n projector  $D$. 

Let us informally recall what is the \calderon projector associated to a elliptic boundary value problem, see Sect. \ref{sec3} for more details:

  let $(N, \hat{g})$ be a complete Riemannian manifold and $P= -\Delta_{\hat{g}}+ m(x)$ for $m\in\cinf(N)$, $m(x)>0$ a Laplace-Beltrami operator. Let also $\Omega\subset M$ a smooth open set.  To $\Omega$ is naturally associated the canonical surface density $dS$, defined by $\langle dS| u\rangle= \int_{\p \Omega}u d\sigma$, for $u\in \coinf(M)$, where $d\sigma$ is the induced surface element on $\p\Omega$.
  
  If $\p_{\nu}$ is the external normal derivative to $\p \Omega$ and $\gamma u= \col{u\tra{\p\Omega}}{\p_{\nu}u\tra{\p \Omega}}$ for $u\in \cinf(\overline{\Omega})$ the \calderon projector $D$ is a map from $\coinf(\p\Omega)\otimes \cc^{2}$ to $\cinf(\p \Omega)\otimes \cc^{2}$ defined by:
 \[
Df\defeq \gamma\circ G(f_{1}(dVol_{\hat{g}})^{-1}dS- f_{0}(dVol_{\hat{g}}^{-1})\p^{*}_{\nu}dS), \ f= \col{f_{0}}{f_{1}}\in \coinf(\p\Omega)\otimes \cc^{2},
 \]
where $G= P^{-1}$. It is easy to see that $f\in \cinf(\Sigma)\otimes \cc^{2}$ equals $\gamma u$ for some $u\in \cinf(\overline{\Omega})$ solution of $Pu=0$ in $\Omega$ if and only if $D f=f$.

In our case we take  $N= \bS_{\beta}\times \Sigma^{+}$, where $\bS_{\beta}$ is the circle of length $\beta$ and $\Sigma^{+}= \Sigma\cap {\mathcal M}^{+}$ is  the right part of the Cauchy surface $\Sigma$. The Riemannian metric  is $\hat{g}= v^{2}(y)d\tau^{2}+ h_{ij}(y)dy^{i}dy^{j}$, obtained by   the {\em Wick rotation}  $ t\eqdef\i \tau$ of the Lorentzian metric $g= - v^{2}(y)dt^{2}+ h_{ij}(y)dy^{i}dy^{j}$ on ${\mathcal M}^{+}\sim \rr\times \Sigma^{+}$ where ${\mathcal M}^{+}$ is identified to $ \rr\times \Sigma^{+}$ using the Killing time coordinate $t$.

The  existence of an extension of $\omega_{\beta_{\rm H}}$ to $M$  is then an almost immediate consequence of the fact that $(N, \hat{g})$ admits  a smooth extension $(N_{\rm ext}, \hat{g}_{\rm ext})$ if  and only if $\beta= \beta_{\rm H}$, a well-known result which plays also a   role in \cite{S}.

 In fact this geometrical fact implies that $D$, viewed as an  operator defined  on $\coinf(\Sigma\backslash \mathcal{B})\otimes \cc^{2}$  uniquely extends to a Calderon projector $D_{\rm ext}$, defined on $\coinf(\Sigma)\otimes \cc^{2}$.  From $D_{\rm ext}$ one can then easily obtain a pure quasi-free state $\omega_{\rm HHI}$ on the whole of $M$.

The Hadamard property of $\omega_{\rm HHI}$ follows then from  the well-known fact that  $D_{\rm ext}$, being  a \calderon projector,  is a  $2\times 2$ matrix of pseudodifferential operators on $\Sigma$, and of the Hadamard property of $\omega_{\beta}$ in ${\mathcal M}^{+}\cup {\mathcal M}^{-}$. 

Beside shortening the proof of the Hadamard property of $\omega_{\rm HHI}$, we think  that our paper 
illustrates the usefulness of pseudodifferential calculus for the construction and study of Hadamard states,  see also \cite{GW1}, \cite{GW2}, \cite{GW3}, \cite{GOW} for other applications.  We believe that \calderon projectors could also be used to construct the Hartle-Hawking-Israel state in the still open case of spacetimes with a Killing horizon that is only stationary.

\subsection{Plan of the paper}
Let us now briefly give the plan of the paper. In Sect. \ref{sec1}  we recall the notion of a static bifurcate Killing horizon, following \cite{S} and introduce the associated Klein-Gordon equation. 

Sect. \ref{secf} is devoted to background material on $CCR^{*}-$algebras, bosonic quasi-free states and their spacetime and Cauchy surface covariances in the case of quantum Klein-Gordon fields. We use the framework of {\em charged fields}, which is in our opinion more elegant, even when considering only neutral field equations. We also recall the notion of {\em pseudodifferential operators} on a manifold, which will be  useful later on and formulate a consequence of  \cite{GW1} which states that the Cauchy surface covariances of {\em any} Hadamard state for Klein-Gordon fields is given by a matrix of pseudodifferential operators.

In Sect. \ref{sec2} we define various 'Euclidean' Laplacians, $K= - \Delta_{\hat{g}}+ m$ acting on $N= \bS_{\beta}\times \Sigma^{+}$ and  a related operator $\tilde{K}$, obtained from Wick rotation of the Lorentzian metric on $M$ in the Killing time coordinate, which are considered in \cite{S}. It is sufficient for us to define these Laplacians by quadratic form techniques, which simplifies some arguments.

In Sect. \ref{sec3} we recall the definition of the {\em Calder\'{o}n projector}, which is a standard notion in elliptic boundary value problems.
In Sect. \ref{sec4},  using the explicit expression for $\tilde{K}^{-1}$, we show that the projection associated to the double $\beta-$KMS state $\omega_{\beta}$ equals to the \calderon projector $D$ associated to $K$ and the open set $\Omega= ]0, \beta/2[\times \Sigma^{+}$.

In Sect. \ref{final},  we recall the well-known fact that a smooth extension $(N_{\rm ext}, \hat{g}_{\rm ext})$ of $(N, \hat{g})$ exists iff $\beta= \beta_{\rm H}$.  The extended \calderon projector $D_{\rm ext}$ generates a pure state on $M$, which is the Hartle-Hawking-Israel state $\omega_{\rm HHI}$. In Prop. \ref{unique}, we show that such an extension is unique among quasi-free states whose spacetime covariances map $\coinf(M)$ into $\cinf(M)$ continuously.
Finally  we give in the proof of Thm. \ref{thmain2} a new and  elementary proof of the Hadamard property of $\omega_{\rm HHI}$, using  the pseudodifferential calculus on $\Sigma$.

\section{spacetimes with a static bifurcate Killing horizon}\label{sec1}\init

\subsection{Static bifurcate Killing horizons}\label{sec1.1}
\begin{figure}[H]
\centering\includegraphics[width=0.5\linewidth]{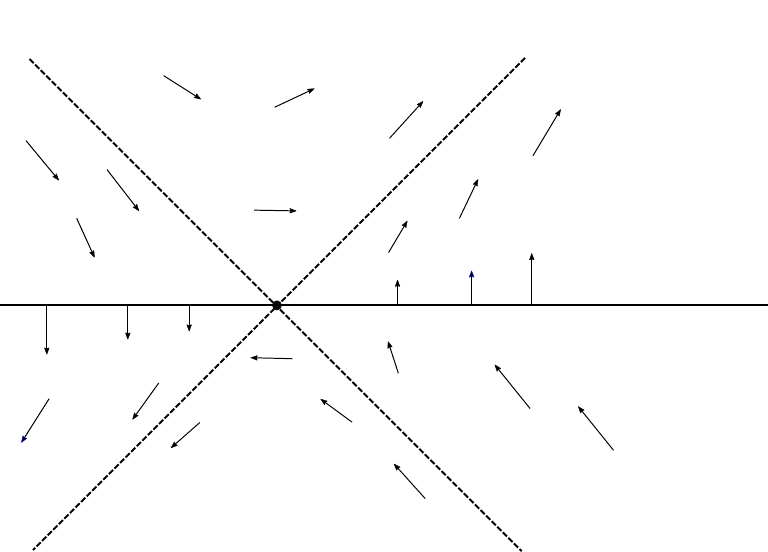}
\put(-10, 50){$\Sigma$ }
\put(-30, 75){${\mathcal M}^{+}$}
\put(-150, 75){${\mathcal M}^{-}$}
\put(-115, 75){$\mathcal{F}$}
\put(-115, 20){$\mathcal{P}$}
\put(-70, 95){$\mathcal{H}^{+}$}
\put(-150, 95){$\mathcal{H}^{-}$}
\put(-65, 10){$\mathcal{H}^{-}$}
\put(-150, 10){$\mathcal{H}^{+}$}
\put(-110, 55){$\mathcal{B}$}
\caption{\label{fig1}}
\end{figure}

We consider as in \cite{S} a globally hyperbolic spacetime $(M, g)$ with a {\em static bifurcate Killing horizon}. We recall, see \cite[Def. 2.2]{S}, that this is a triple $(M, g, V)$, such that
\ben
\item   the Lorentzian manifold $(M, g)$ is globally hyperbolic,
\item   $V$ is a complete Killing vector field for $(M, g)$,
\item $\cB\defeq\{x\in M : V(x)=0\}$ is an  compact, orientable submanifold of codimension $2$,
\item there exists a Cauchy hypersurface $\Sigma$ containing $\cB$,
\item $V$ is  $g-$orthogonal to $\Sigma$,
\een
see  Figure \ref{fig1} below  where the vector field $V$ is represented by arrows.
For simplicity we will also assume that  the bifurcation surface $\cB$ is connected. Denoting by $n$ the future pointing normal vector field to $\Sigma$ one introduces the {\em lapse function}:
\begin{equation}
\label{e0.1}
v(x)\defeq - n(x)\dito g(x)V(x), \ x\in \Sigma,
\end{equation} and  $\Sigma$ decomposes as 
\[
\Sigma= \Sigma^{-}\cup \cB\cup \Sigma^{+},
\]
where $\Sigma^{\pm}=\{x\in\Sigma: \pm v(x)>0\}$. The spacetime $M$ splits as 
\[
M= {\mathcal M}^{+}\cup {\mathcal M}^{-}\cup \overline{\mathcal{F}}\cup\overline{\mathcal{P}},
\]
where  the future cone $\mathcal{F}\defeq I^{+}(\cB)$, the past cone $\mathcal{P}\defeq I^{-}(\cB)$, the right/left wedges ${\mathcal M}^{\pm}\defeq D(\Sigma^{\pm})$, are all globally hyperbolic when equipped  with  $g$.
 \subsection{Wedge reflection}\label{sec1.2}
Additionally one has to assume the existence of a {\em wedge reflection}, see \cite[Def. 2.6]{S}, ie a diffeomorphism $R$ of ${\mathcal M}^{+}\cup {\mathcal M}^{-}\cup U$ onto itself, where $U$ is an open neighborhood of $\cB$ such that:
\ben
\item $R\circ R = {\rm Id}$,
\item $R$ is an isometry of $({\mathcal M}^{+}\cup {\mathcal M}^{-}, g)$ onto itself, which reverses the time orientation,
\item $R= {\rm Id}$ on $\cB$,
\item $R^{*}V= V$ on ${\mathcal M}^{+}\cup {\mathcal M}^{-}$.
\een
It follows that $R$ preserves $\Sigma$, see \cite[Prop. 2.7]{S}, and we denote by $r$ the restriction of $R$ to $\Sigma$. Denoting by  $h$ the induced Riemannian metric on $\Sigma$ one has:
\beq\label{proptoc}
r^{*}h= h, r^{*}v= -v.
\eeq
\subsection{Killing time coordinate}\label{sec1.4}
Denoting by $\Phi^{V}_{s}: M\to M$ the flow of the Killing vector field $V$, we obtain a diffeomorphism
\[
\chi: \rr\times (\Sigma\backslash \cB)\ni (t,y)\mapsto \Phi_{t}^{V}(y)\in {\mathcal M}^{+}\cup {\mathcal M}^{-},
\]
which defines the coordinate $t$ on ${\mathcal M}^{+}\cup {\mathcal M}^{-}$ called the {\em Killing time coordinate}. The metric $g$ on ${\mathcal M}^{+}\cup {\mathcal M}^{-}$ pulled back by $\chi$ takes the form (see \cite[Subsect. 2.1]{S}):
\begin{equation}
\label{e0.4}
g= - v^{2}(y)dt^{2}+ h_{ij}(y)dy^{i}dy^{j},
\end{equation}
where  the Riemannian metric $ h_{ij}(y)dy^{i}dy^{j}$ is the restriction of $g$ to $\Sigma$.
\subsection{Klein-Gordon operator}\label{sec1.3}
We fix a real potential $m\in \cinf(M)$. As in \cite{S} we assume that $m$ is stationary w.r.t. the Killing vector field $V$ and invariant under the wedge reflection, ie:
\begin{equation}
\label{e0.2}
V^{a}\nabla_{a}m(x)=0, \ m\circ R(x)= m(x), \ x\in {\mathcal M}^{+}\cup {\mathcal M}^{-}\cup U. 
\end{equation}
For simplicity we also assume that 
\begin{equation}
\label{e0.3}
m(x)\geq m_{0}^{2}>0, \ x\in M,
\end{equation}
ie  we consider only massive fields.
Note that in \cite{S} the weaker condition $m(x)>0$ was assumed.
We consider the Klein-Gordon operator
\begin{equation}
\label{e0.4b}
P= - \Box_{g}+ m.
\end{equation}

\section{Free Klein-Gordon fields}\label{secf}\init
In this section we briefly recall some well-known background material on free quantum Klein-Gordon fields on globally hyperbolic spacetimes.  We follow the presentation in \cite[Sect. 2]{GW1} based on {\em charged fields}.
\subsection{Charged CCR algebra}\label{secf.1}
\subsubsection{Charged bosonic fields}
Let $\cY$ a complex vector space, $\cY^{*}$ its anti-dual. Sesquilinear forms on $\cY$ are identified with elements of $L(\cY, \cY^{*})$ and the action of a sesquilinear form $\beta$ is correspondingly denoted by $\bar{y}_{1}\dito \beta y_{2}$ for $y_{1}, y_{2}\in \cY$. We fix  $q\in L_{\rm h}(\cY, \cY^{*})$  a non degenerate hermitian form on $\cY$, ie such that $\Ker q=\{0\}$.

  The {\em CCR} $^{*}-${\em algebra} $\CCR(\cY, q)$ is the complex $^{*}-$algebra generated by symbols $\one, \psi(y), \psi^{*}(y),y\in \cY$ and the relations:
\[
\begin{array}{l}
\psi(y_{1}+ \lambda y_{2})= \psi(y_{1})+ \bar{\lambda}\psi(y_{2}), \ y_{1}, y_{2}\in \cY, \lambda\in \cc,\\[2mm]
\psi^{*}(y_{1}+ \lambda y_{2})= \psi^{*}(y_{1})+ \lambda\psi^{*}(y_{2}), \ y_{1}, y_{2}\in \cY, \lambda\in \cc,\\[2mm]
[\psi(y_{1}), \psi(y_{2}]= [\psi^{*}(y_{1}), \psi^{*}(y_{2})]=0, \ [\psi(y_{1}), \psi^{*}(y_{2})]= \bar{y}_{1}\dito qy_{2}\one, \ y_{1}, y_{2}\in\cY,\\[2mm]
\psi(y)^{*}= \psi^{*}(y), \ y\in \cY.
\end{array}
\]
 A state $\omega$ on $\CCR(\cY, q)$ is {\em (gauge invariant) quasi-free} if 
 \[
\omega(\prod_{i=1}^{p}\psi(y_{i})\prod_{i=1}^{q}\psi^{*}(y_{j}))=\left\{  \begin{array}{l}
0\hbox{ if }p\neq q,\\
\sum_{\sigma\in S_{p}}\prod_{i=1}^{p}\omega(\psi(y_{i})\psi^{*}(y_{\sigma(i)}))\hbox{ if }p=q.
\end{array} \right.
\]
There is no loss of generality to restrict oneself to charged fields and gauge invariant states, see eg the discussion in \cite[Sect. 2]{GW1}. 
It is convenient to associate to  $\omega$ its {\em (complex) covariances} $\lambda^{\pm}\in L_{\rm h}(\cY, \cY^{*})$ defined by:
\[
\begin{array}{l}
\omega(\psi(y_{1})\psi^{*}(y_{2}))\eqdef \bar{y}_{1}\dito \lambda^{+} y_{2}, \\[2mm]
 \omega(\psi^{*}(y_{2})\psi(y_{1}))\eqdef \bar{y}_{1}\dito \lambda^{-} y_{2},
 \end{array} \ y_{1}, y_{2}\in \cY.
\]
The following results are well-known, see eg \cite[Sect. 17.1]{DG} or \cite[Sect. 2]{GW1}:

  -- two  hermitian  forms $\lambda^{\pm}\in L_{\rm h}(\cY, \cY^{*})$ are the covariances of a quasi-free state $\omega$ iff
 \begin{equation}
\label{ef.1}
\lambda^{\pm}\geq 0, \ \lambda^{+}- \lambda^{-}=q.
\end{equation}
 -- Let  $\cY_{\omega}$ be the completion of $\cY$ for the Hilbertian scalar product $\lambda^{+}+ \lambda^{-}$. If there exist linear operators $c^{\pm}\in L(\cY_{\omega})$ such that 
\[
c^{+}+ c^{-}= \one, \ (c^{\pm})^{2}= c^{\pm},
\]
(ie   $c^{\pm}$ is a pair of complementary projections) and  $\lambda^{\pm}= \pm q\circ c^{\pm}$, then $\omega$ is a {\em pure state}.
\subsubsection{Neutral bosonic fields}
We complete this subsection by explaining the relationship with the formalism of neutral fields, see eg \cite[Subsect. 2.5]{GW1}. 

Let $\cX$  a real vector space, $\cX^{\t}$ its dual, and $\sigma\in L_{\a}(\cX, \cX^{\t})$  a symplectic form on $\cX$. The $^{*}-$algebra $\CCR(\cX, \sigma)$ is the complex  $^{*}-$algebra generated by symbols $\one, \phi(x), x\in \cX$ and relations:
\[
\begin{array}{l}
\phi(x_{1}+ \lambda x_{2})= \phi(x_{1})+ \lambda\phi(x_{2}), \ x_{1}, x_{2}\in \cX, \lambda\in \rr,\\[2mm]
[\phi(x_{1}), \phi(x_{2}]= \i  x_{1}\dito \sigma x_{2}\one, \ x_{1}, x_{2}\in\cX,\\[2mm]
\phi(x)^{*}= \phi(x), \ x\in \cX.
\end{array}
\]
To relate the neutral to the charged formalism one sets
 $\cY= \cc\cX$ and  for $\beta\in L(\cX, \cX^{\t})$ denote by $\beta_{\cc}\in L(\cY, \cY^{*})$ its sesquilinear extension.    $\cY_{\rr}\sim \cX\oplus \cX$  is the {\em real form} of $\cY$, ie $\cY_{\rr}= \cY$ as a real vector space.   Then   $(\cY_{\rr}, {\rm Re}\sigma_{\cc})\sim (\cX, \sigma)\oplus (\cX, \sigma)$ is a real symplectic space and we denote by $\phi(y), y\in \cY_{\rr}$ the selfadjoint generators of $\CCR(\cY_{\rr}, {\rm Re}\sigma_{\cc})$.
 Under the identification$\phi(y)\sim \phi(x)\otimes\one + \one\otimes \phi(x')$ for $y= x+ \i x'$ we can identify $\CCR(\cY_{\rr}, {\rm Re}\sigma_{\cc})$ with $\CCR(\cX, \sigma)\otimes \CCR(\cX, \sigma)$ as $^{*}-$algebras.
 
 Note also that under the identification
 \[
\psi(y) \sim \frac{1}{\sqrt{2}}(\phi(y)+\i \phi(i y)), \ \psi^{*}(y)\sim  \frac{1}{\sqrt{2}}(\phi(y)-\i \phi(i y)), \ y\in \cY
\]
we can identify $\CCR(\cY_{\rr}, {\rm Re}\sigma_{\cc})$ with $\CCR(\cY, q)$ for $q= \i \sigma_{\cc}$.

 A quasi-free state $\omega$ on $\CCR(\cX, \sigma)$ is determined by its {\em real covariance} $\eta\in L_{\rm s}(\cX, \cX^{\t})$ defined by:
 \[
 \omega(\phi(x_{1})\phi(x_{2}))\eqdef x_{1}\dito \eta x_{2}+ \frac{\i}{2}x_{1}\dito \sigma x_{2}, \ x_{1}, x_{2}\in \cX.
 \]
 A symmetric form $\eta\in L_{\rm s}(\cX, \cX^{\t})$ is the covariance of a quasi-free state iff 
 \[
  \eta\geq 0, \ |x_{1}\dito \sigma x_{2}|\leq 2 (x_{1}\dito \eta x_{1})^{\12} (x_{2}\dito  \eta x_{2})^{\12}, \ x_{1}, x_{2}\in \cX.
 \]
 To such a state $\omega$ we associate the quasi-free state $\tilde{\omega}$ on $\CCR(\cY_{\rr}, {\rm Re}\sigma_{\cc})$ with real covariance ${\rm Re}\eta_{\cc}$.  Then its complex covariances $\lambda^{\pm}$ are given by (see \cite[Subsect. 2.5]{GW1}):
 \begin{equation}
\label{ef.2}
\lambda^{\pm}= \eta_{\cc}\pm \12 \i\sigma_{\cc}.
\end{equation}
Applying complex conjugation, we immediately see that  in this case
\begin{equation}
\label{ef.2bb} \lambda^{+}\geq 0 \ \Leftrightarrow \ \lambda^{-}\geq 0,
\end{equation}
so it suffices to check for example that $\lambda^{+}\geq 0$.
\subsection{Free Klein-Gordon fields}\label{secf2}
Let $P= -\Box_{g}+ m(x)$, $m\in \cinf(M, \rr)$ a Klein-Gordon operator on a globally hyperbolic spacetime $(M, g)$ (we use the convention $(1,d)$ for the Lorentzian signature). Let  $E^{\pm}$  be the advanced/retarded inverses of $P$ and  $E\defeq E^{+}- E^{-}$.   We apply the above framework to
\[
\cY= \frac{\coinf(M)}{P\coinf(M)}, \ \overline{[u]}\dito q[u]= \i (u| Eu)_{M},
\]
where $(u|v)_{M}= \int_{M}\bar{u}v dVol_{g}$.

One restricts attention to quasi-free states on $\CCR(\cY, q)$ whose covariances are given by distributions on $M\times M$, ie such that there exists $\Lambda^{\pm}\in \cD'(M\times M)$ with
\beq\label{ef.2b}
\begin{array}{l}
\omega(\psi([u_{1}])\psi^{*}([u_{2}]))= (u_{1}| \Lambda^{+}u_{2})_{M}, \\[2mm]
 \omega(\psi^{*}([u_{2}])\psi([u_{1}]))= (u_{1}| \Lambda^{-}u_{2})_{M},
\end{array}
 \ u_{1}, u_{2}\in \coinf(M).
\eeq
In the sequel the distributions $\Lambda^{\pm}\in \cD'(M\times M)$ will be called the {\em spacetime covariances} of  the state $\omega$.

In \eqref{ef.2b} we identify distributions on $M$ with distributional densities using the density $dVol_{g}$ and use the notation $(u|\varphi)_{M}$, $u\in \coinf(M)$, $\varphi\in \cD'(M)$ for the duality bracket. We have then
\beq\label{tralala}
\begin{array}{l}
P(x, \p_{x})\Lambda^{\pm}(x, x')= P(x', \p_{x'})\Lambda^{\pm}(x, x')=0,\\[2mm]
 \Lambda^{+}(x, x')- \Lambda^{-}(x, x')= \i E(x, x').
\end{array}
\eeq
 Such a state is called a {\em Hadamard state} , (see \cite{R} for the neutral case and \cite{GW1} for the complex case) if 
 \def\WF{{\rm WF}}\def\cN{\mathcal{N}}
 \begin{equation}
\label{ef.3}
\WF(\Lambda^{\pm})'\subset \cN^{\pm}\times \cN^{\pm},
\end{equation}
where $\WF(\Lambda)'$ denotes the 'primed' wavefront set of $\Lambda$, ie $S'\defeq\{((x, \xi), (x', -\xi')): ((x, \xi), (x', \xi'))\in S\}$ for $S\subset T^{*}M\times T^{*}M$, and $\cN^{\pm}$ are the two connected components (positive/negative energy shell) of  the characteristic manifold:
\beq\label{defdechar}
\cN\defeq \{(x, \xi)\in T^{*}M\backslash\{0\}):  \xi_{\mu}g^{\mu\nu}(x)\xi_{\nu}=0\}.
\eeq
\subsection{Cauchy surface covariances}\label{csc}
Denoting by $Sol_{\rm sc}(P)$ the space of smooth space-compact solutions of $P\phi=0$, it is well known that 
\[
[E]:\frac{\coinf(M)}{P\coinf(M)}\ni [u]\mapsto Eu\in   Sol_{\rm sc}(P)
\]
is bijective, with 
\[
\i (u_{1}| Eu_{2})= \overline{Eu}_{1}\dito q Eu_{2}, \ u_{i}\in \coinf(M),
\]
for 
\beq\label{defdeq}
\overline{\phi}_{1}\dito q \phi_{2}\defeq\i\int_{\Sigma}(\nabla_{\mu}\bar{\phi}_{1}\phi_{2}- \bar{\phi}_{1}\nabla_{\mu}\phi_{2})n^{\mu}d\sigma,
\eeq
where $\Sigma$ is any spacelike Cauchy hypersurface, $n^{\mu}$ is the future directed unit normal vector field to $\Sigma$ and $d\sigma$ the induced surface density.  Setting
\[
\rho: \cinf_{\rm sc}(M)\ni \phi\mapsto \col{\phi\tra\Sigma}{\i^{-1}\p_{\nu}\phi\tra\Sigma}= f\in \coinf(\Sigma)\oplus \coinf(\Sigma)
\]
Since the  Cauchy problem 
\[
\left\{
\begin{array}{rl}
P\phi=0, \\
\rho u=f
\end{array}
\right.
\]
as a unique solution $\phi\in Sol_{\rm sc}(P)$ for $f\in \coinf(\Sigma)\oplus \coinf(\Sigma)$ the map
\[
\frac{\coinf(M)}{P\coinf(M)}\ni [u]\mapsto \rho Eu\in \coinf(\Sigma)\oplus \coinf(\Sigma)
\]
is bijective, and
\[
\i (u|Eu)_{M}= \overline{\rho E u}\dito q \rho Eu, 
\]
for
\begin{equation}
\label{ef.3b}
\bar{f}\dito q f\defeq\int_{\Sigma} \bar{f}_{1} f_{0}+ \bar{f}_{0}f_{1}d \sigma_{\Sigma}, \ f= \col{f_{0}}{f_{1}}.
\end{equation}
 It follows that   to a quasi-free state with spacetime covariances $\Lambda^{\pm}$ one can associate its {\em Cauchy surface covariances} $\lambda^{\pm}$ defined by:
 \begin{equation}
\label{ef.4}
\Lambda^{\pm}\eqdef (\rho E)^{*} \lambda^{\pm}( \rho E).
\end{equation}
 Using the canonical scalar product  $(f|f)_{\Sigma}\defeq\int_{\Sigma} \bar{f}_{1}f_{1}+ \bar{f}_{0}f_{0}d\sigma_{\Sigma}$ we identify $\lambda^{\pm}$ with operators, still denoted by $\lambda^{\pm}$, belonging to $L(\coinf(\Sigma)\oplus \coinf(\Sigma), \cD'(\Sigma)\oplus \cD'(\Sigma))$.
 
 A more explicit expression of $\lambda^{\pm}$ in terms of $\Lambda^{\pm}$ is as follows, see eg \cite[Thm. 7.10]{GOW}: 
 let us introduce Gaussian normal coordinates to $\Sigma$ 
\[
U\ni (t, y)\mapsto \chi(t,y)\in V,
\]
where $U$ is an open  neighborhood of $\{0\}\times \Sigma$ in $\rr\times \Sigma$ and $V$ an open neighborhood of $\Sigma$ in $M$, such that $\chi^{*}g= - dt^{2}+ h_{ij}(t,y)dy^{i}dy^{j}$.  We denote by $\Lambda^{\pm}(t, y, t',y' )\in \cD'(U\times U)$ the restriction to $U\times U$ of the distributional kernel of $\Lambda^{\pm}$.   By \eqref{tralala} and standard microlocal arguments, their restrictions to fixed times $t,t'$, denoted by $\Lambda^{\pm}(t, t')\in \cD'(\Sigma\otimes \Sigma)$ are well defined.

We know also that $\p^{k}_{t}\p^{k'}_{t'}\Lambda^{\pm}(0, 0)\in \cD'(\Sigma\times \Sigma)$ is well defined for $k, k'=0, 1$.  Then setting $\lambda^{\pm}\eqdef \pm q\circ c^{\pm}$ we have:
\beq\label{defdecplusmoins}
c^{\pm}= \pm\mat{\i \p_{t'}\Lambda^{\pm}(0, 0)}{\Lambda^{\pm}(0, 0)}{\p_{t} \p_{t'}\Lambda^{\pm}(0, 0)}{\i^{-1} \p_{t}\Lambda^{\pm}(0, 0)}.
\eeq

Large classes of Hadamard states were constructed in terms of their Cauchy surface covariances in \cite{GW1, GOW} using pseudodifferential calculus on $\Sigma$, see below for a short summary.
\subsection{Pseudodifferential operators}\label{sec3.1}
We briefly recall the notion of (classical) pseudodifferential operators on a manifold, referring to \cite[Sect. 4.3]{Sh} for details.

For $m\in \rr$ we denote by $\Psi^{m}(\rr^{d})$ the space of classical pseudodifferential operators of order $m$ on $\rr^{d}$, associated with poly-homogeneous symbols of order $m$ see eg \cite[Sect. 3.7]{Sh}. 

Let $N$ be a smooth, $d-$di\-mensional   manifold.  Let $U\subset N$ a precompact   chart open set and $\psi: U\to \tilde{U}$ a chart diffeomorphism,  where $\tilde{U}\subset \rr^{d}$ is precompact, open.  We denote by $\psi^{*}: \coinf(\tilde{U})\to \coinf(U)$ the map $\psi^{*} u(x)\defeq u\circ \psi(x)$.\begin{definition}
 A linear continuous map $A: \coinf(N)\to \cinf(N)$ belongs to $\Psi^{m}(N)$ if  the following condition holds:
 
(C) Let    $U\subset N$    be precompact open, $\psi: U\to \tilde{U}$ a chart diffeomorphism, $\chi_{1}, \chi_{2}\in \coinf(U)$ and $\tilde{\chi}_{i}= \chi_{i}\circ \psi^{-1}$. Then there exists $\tilde{A}\in \Psi^{m}(\rr^{d})$ such that
 \beq\label{eapp.-4}
(\psi^{*})^{-1} \chi_{1}A \chi_{2}\psi^{*}= \tilde{\chi}_{1}\tilde{A}\tilde{\chi}_{2}.
\eeq
 Elements of $\Psi^{m}(N)$ are called {\em (classical) pseudodifferential operators} of order $m$ on $N$. 
 
 The subspace of $\Psi^{m}(N)$ of pseudodifferential operators with {\em properly supported kernels} is denoted by $\Psi^{m}_{\rm c}(N)$.
\end{definition}
Note that  if $\Psi^{\infty}_{(\rm c)}(N)\defeq \bigcup_{m\in \rr}\Psi^{m}_{(\rm c)}(N)$, then $\Psi^{\infty}_{\rm c}(N)$ is an algebra, but $\Psi^{\infty}(N)$ is not, since without the proper support condition, pseudodifferential operators cannot in general be composed.  

We denote by $T^{*}N\backslash\{0\}$ the cotangent bundle of $N$ with the zero section removed.

To $A\in \Psi^{m}(N)$ one can associate its {\em principal symbol} $\sigma_{\rm pr}(A)\in \cinf(T^{*}N\backslash\{0\})$, which is homogeneous of degree $m$ in the  fiber variable $\xi$ in $T^{*}M$, in $\{|\xi|\geq 1\}$. $A$ is called {\em elliptic}  in $\Psi^{m}(N)$ at $(x_{0}, \xi_{0})\in T^{*}N\backslash\{0\}$ if $\sigma_{\rm pr}(A)(x_{0}, \xi_{0})\neq 0$.

If $A\in \Psi^{m}(N)$ there exists (many) $A_{\rm c}\in \Psi^{m}_{\rm c}(N)$ such that $A-A_{\rm c}$ has a smooth kernel.

Finally one says that $(x_{0}, \xi_{0})\not\in {\rm essupp}(A)$ for $A\in \Psi^{\infty}(N)$ if there exists $B\in \Psi^{\infty}_{\rm c}(N)$ {\em elliptic} at $(x_{0}, \xi_{0})$ such that $A_{\rm c}\circ B$ is smoothing, where $A_{\rm c}\in \Psi^{\infty}_{\rm c}(N)$ is as above, ie $A- A_{\rm c}$ is smoothing.

\subsection{The Cauchy surface covariances of Hadamard states}
We now state a result which follows directly from a construction of Hadamard states in \cite[Subsect. 8.2]{GW1}.
\begin{theorem}\label{allhad}
 Let $\omega$ be any Hadamard state for the free Klein-Gordon field on $(M, g)$ and $\Sigma$ a spacelike Cauchy hypersurface. Then its Cauchy surface covariances $\lambda^{\pm}$ are $2\times 2$ matrices  with entries in $\Psi^{\infty}(\Sigma)$. 
\end{theorem}
\proof 
It is well known (see eg \cite{R}) that  if $\omega_{1}, \omega_{2}$ are Hadamard states, then $\Lambda^{\pm}_{1}- \Lambda^{\pm}_{2}$ are smoothing operators on $M$. Using \eqref{ef.4} this implies that $\lambda^{\pm}_{1}- \lambda^{\pm}_{2}$ are matrices of smoothing operators on $\Sigma$. From the definition of $\Psi^{\infty}(\Sigma)$ it hence suffices to construct {\em one} Hadamard state $\omega$ whose Cauchy surface covariances $\lambda^{\pm}$
 are matrices of pseudodifferential operators.  The state constructed in \cite[Subsect. 8.2]{GW1} has this property, as  can be seen from \cite[Equ. (8.2)]{GW1}. \qed

\section{Euclidean operators}\label{sec2}\init
The construction of the $\beta-$KMS state on ${\mathcal M}^{+}$ with respect to the Killing vector field $V$ relies on the  {\em Wick rotation}, where  $(\rr\times \Sigma^{+}, g)$  is replaced by $(\bS_{\beta}\times \Sigma^{+}, \hat{g})$: 
\beq\label{trof}
\hat{g}= v^{2}(y)d\tau^{2}+ h_{ij}(y)dy^{i}dy^{j},
\eeq
is the Riemannian metric obtained from \eqref{e0.4} by setting $t= \i \tau$ and  $\bS_{\beta}= [0,\beta[$ with endpoints identified  is the circle of length $\beta$.

In this section we recall various 'Euclidean' operators related to $\hat{g}$ appearing in \cite{S, S2}.  It will be convenient to construct them by quadratic form techniques.

We set
\[
N\defeq \bS_{\beta}\times \Sigma^{+},
\] 
whose elements are denoted by $(\tau, y)$. We equip $N$ with the Riemannian metric $\hat{g}$ in \eqref{trof} and the associated density $dVol_{\hat{g}}= |v|(y)|h|^{\12}(y)d\tau dy$. The hypersurface $\Sigma^{+}$ is equipped with the induced density $dVol_{h}= |h|^{\12}(y)dy$.

\subsection{Euclidean operator on $N$}\label{sec2.1}
We consider the operator 
\[
K\defeq - \Delta_{\hat{g}}+ m(y),
\]
for $m$ as in Subsect. \ref{sec1.3}. Note that $m$ depends only on $y$ since $m$ is invariant under the Killing flow.  We have
\[
K= - v^{-2}(y)\pe_{\tau}^{2}- |v|^{-1}(y)|h|^{-\12}(y)\pe_{y^{i}}|v|(y)|h|^{\12}(y)h^{ij}(y)\pe_{y^{j}}+ m(y).
\]
$K$ is well defined as a selfadjoint operator on $L^{2}(N, dVol_{\hat{g}})$ obtained from the quadratic form:
\beq\label{defdequad}
Q(u, u)\defeq \int_{N}\left(|v|^{-2}|\p_{\tau}u|^{2}+ \p_{i}\bar{u}h^{ij}\p_{j}u + m|u|^{2}\right)dVol_{\hat g},
\eeq
which is closeable on $\coinf(N)$, since   $K$ is symmetric and bounded from below on this domain. Denoting its closure again by $Q$ and the domain of its closure by $\Dom Q$, $K$ is the selfadjoint operator associated to $Q$, ie the Friedrichs extension of  its restriction to $\cinf(\bS_{\beta})\otimes \coinf(\Sigma^{+})$.
We know that $u\in \Dom K$, $Ku= f$ iff 
\begin{equation}
\label{e1.1d}
u\in \Dom Q\hbox{ and }Q(w, u)= (w|f)_{L^{2}(N)}, \ \forall w\in \coinf(N).
\end{equation}

From \eqref{e0.3} we know that $K\geq m_{0}^{2}$ hence is boundedly invertible and we set
 \[
G\defeq K^{-1}.
\]
\subsection{Change of volume form}\label{sec2.2}
Let us set $\hat{Q}(u, u)= Q(vu, vu)$, $\Dom \hat{Q}= \{u\in L^{2}(N): vu\in \Dom Q\}$. By \eqref{e0.3} we have $\hat{Q}(u, u)\geq m_{0}^{2}\| vu\|^{2}$. If $u_{n}\in \Dom \hat{Q}, u\in L^{2}(N)$ with $\| u_{n}-u\|\to 0$ and $\hat{Q}(u_{n}- u_{m}, u_{n}- u_{m})\to 0$ then from the inequality above we obtain that $vu\in L^{2}(N)$ and $\| v(u_{n}- u)\|\to 0$. Since $Q$ is closed we obtain that $u\in \Dom \hat{Q}$ and $\hat{Q}(u_{n}- u, u_{n}-u)\to 0$, ie $\hat{Q}$ is closed. 

Let $\hat{K}$ be the injective selfadjoint operator associated to $\hat{Q}$, (which is formally equal to $vK v$) and let $\hat{G}= \hat{K}^{-1}$.    We claim that
\begin{equation}
\label{e2.toti}
G= v\hat{G}v, \hbox{ on }v^{-1}L^{2}(N).
\end{equation}
This follows  easily from the caracterization \eqref{e1.1d} of $G$ and similarly of $\hat{G}$. 

Let now $U: L^{2}(N)\to  L^{2}(\bS_{\beta})\otimes L^{2}(\Sigma^{+})$ the unitary map given by $Uu= v^{\12}u$. We set
\[
\tilde{K}\defeq U \hat{K}U^{*}.
\]
  We have 
\[
\tilde{K}=-\pe_{\tau}^{2}+ \epsilon^{2}(y, \pe_{y}),
\]
where:
\[
\epsilon^{2}(y, \pe_{y})= - |v|^{\12}(y)|h|^{-\12}(y)\pe_{y^{i}}|v|(y)|h|^{\12}(y)h^{ij}(y)\pe_{y^{j}}|v|^{\12}(y)+ v^{2}(y)m(y),
\]
is obtained  as above  from the quadratic form
\beq\label{e2.tito}
\int_{\Sigma^{+}}\left(\p_{i}|v|^{\12}\overline{\tilde{u}}|v|h^{ij}\p_{i}|v|^{\12}u+ |v|^{2}m|u|^{2}\right)|h|^{\12}dy.
\eeq
 If $\tilde{G}\defeq \tilde{K}^{-1}$ we have by \eqref{e2.toti}:
\begin{equation}
\label{e1.2}
G=  |v|^{1/2}\tilde{G}|v|^{3/2}, \hbox{ on }v^{-3/2}L^{2}(N).
\end{equation}
 We now recall a well known expression for $\tilde{G}$.
Let
\[
F(\tau)= \dfrac{\e^{- \tau \epn}+ \e^{(\tau- \beta)\epn}}{2\epn(1- \e^{- \beta \epn})},\ \tau\in [0, \beta[,
\]
extended to $\tau\in\rr$ by $\beta-$periodicity.  In particular we have:
\begin{equation}
\label{e1.3}
F(\tau)=  \dfrac{\e^{- |\tau| \epn}+ \e^{(|\tau|- \beta)\epn}}{2\epn(1- \e^{- \beta \epn})},\ \tau\in [-\beta,\beta]
\end{equation}
The following expression for $\tilde{G}$ is well-known (see eg \cite[Sect. 18.3.2]{DG}):
\begin{equation}
\label{e1.4}
\tilde{G}\tilde{u}(\tau)= \int_{\bS_{\beta}} F(\tau-\tau')\tilde{u}(\tau')d\tau', \ \tilde{u}\in L^{2}(S_{\beta})\otimes L^{2}(\Sigma\backslash \cB).
\end{equation}
Note that since $\epsilon^{2}\geq mv^{2}$ by \eqref{e2.tito}, we have also $\epsilon^{-2}\leq m^{-1}v^{-2}$ by Kato-Heinz theorem hence 
$\coinf\Sigma^{+})\subset \Dom F(\tau)$.

\section{Calder\'{o}n projectors}\label{sec3}\init
In this section we recall some standard facts on  {\em Calder\'{o}n projectors}. We refer the reader to \cite[Sects. 5.1- 5.3]{CP} for details.
\subsection{The Calder\'{o}n projector}\label{sec3.2}
Let $(N, h)$ a complete Riemannian manifold and $P= - \Delta_{h}+ m$, where $m\in \cinf(N)$ is a real potential with $m(x)\geq m_{0}^{2}>0$. 
 As in Sect. \ref{sec2} we construct $P$  as a selfadjoint operator on $L^{2}(N, dVol_{h})$  using the quadratic form 
 \beq\label{quado}
 Q(u, u)= \int_{N}\p_{i}\bar{u}h^{ij}\p_{j}u+ m(x)|u|^{2}(x)dVol_{h}.
 \eeq
   We obtain that  $0\in \rho(P)$, hence $G\defeq P^{-1}$ is  a bounded operator on $L^{2}(N, dVol_{h})$,  defined by
 \begin{equation}
\label{e3.10}
Q(Gv, w)= (v|w)_{L^{2}(N)}, \ \forall w\in \coinf(N).
\end{equation}

 Let $\Omega\subset N$ an open set such that $\p \Omega= S= \bigcup_{1}^{n}S_{i}$, where  
 $S_{i}$ are the connected components  of $S$ and are assumed to be   smooth hypersurfaces.  We denote by $\cinf(\overline{\Omega})$ the space of restrictions to $\Omega$ of functions in $\cinf(N)$.
 
  We  associate to $S_{i}$ the distribution density $dS_{i}$ defined by:
  \[
\langle dS_{i}| u\rangle\defeq \int_{S_{i}} ud\sigma^{(i)}_{h}, \ u\in\coinf(N),
\]
where $d\sigma^{(i)}_{h}$ is the induced Riemannian density on $S_{i}$ and we set
\[
dS= \sum_{i=1}^{n}dS_{i}.
\]
We denote by $\p_{\nu}$ the unit exterior normal vector field to $S$ and set
\[
\langle \p_{\nu}^{*}dS| u\rangle\defeq \langle dS| \p_{\nu}u\rangle, \ u\in \coinf(N).
\]

 For $u\in \cinf(\overline{\Omega})$ we set
 \[
\gamma u\defeq \col{u\tra{S}}{\p_{\nu}u\tra{S}}=: \col{\gamma_{0}u}{\gamma_{1}u}.
\]
For $v\in \coinf(S)$ we denote by $\tilde{v}\in \coinf(N)$ an extension of $v$ to $N$ such that $\tilde{u}\tra{S}=u, \p_{\nu}\tilde{u}\tra{S}=0$.
 \begin{definition}
 Let $f= \col{f_{0}}{f_{1}}\in \coinf(S)\oplus\coinf(S)$. We set:
 \[
Df\defeq \gamma\circ G(\tilde{f}_{1}(dVol_{h})^{-1}dS- \tilde{f}_{0}(dVol_{h}^{-1})\p^{*}_{\nu}dS).
\]
-- The operator $D:\coinf(S)\oplus \coinf(S)\to \cinf(S)\oplus \cinf(S)$ is continuous and is called the {\em Calder\'{o}n projector} associated to $(P, S)$.

-- The operator $D$ is a $2\times 2$ matrix of pseudodifferential operators on $S$.
\end{definition}
Note that $dS$ and $\p_{\nu}^{*}dS$ are distributional densities, hence $(dVol_{h})^{-1}dS$ and $(dVol_{h})^{-1}\p_{\nu}^{*}dS$ are distributions on $N$, supported on $S$.

Note also that  the Calder\'{o}n projector is obviously covariant under diffeomorphisms: if $\chi: (N, h)\to (N', h')$ is an isometric diffeomorphism with $S'= \chi(S)$, $P= \chi^{*}P'$, then   
\[
D= \psi^{*}D',
\]
 where $\psi: S\to S'$ is the restriction of $\chi$ to $S$.
 
 \subsubsection{Expression in Gaussian normal coordinates}
Let $U_{i}$ be a neighborhood of $\{0\}\times S_{i}$ in $\rr\times S_{i}$  and $V_{i}$ a neighborhood of $S_{i}$ in $N$ such that Gaussian normal coordinates to $S_{i}$ induce a diffeomorphism:
\[
\chi_{i}: U_{i}\ni x\mapsto (s, y)\in V_{i}
\]
 from $U_{i}$ to $V_{i}$,  and $ds^{2}+ k_{s}(y)dy^{2}= \chi_{i}^{*}h$  on $U_{i}$. Then  for $f\in \coinf(S_{i})\otimes \cc^{2}$ we have
\beq\label{defdet}
\begin{array}{rl}
&\chi_{i}^{*}\left(\tilde{f}_{1}(dVol_{h})^{-1}dS -\tilde{f}_{0}(dVol_{h})^{-1}\p_{\nu}^{*}dS\right)\\[2mm]
=& \delta_{0}(s)\otimes (f_{1}(y)-r_{0}(y)f_{0}(y))- \delta_{0}'(s)\otimes f_{0}(y),
\end{array}
\eeq
where $r_{s}(y)= |k_{s}|^{-\12}(y)\p_{s}|k_{s}|^{\12}$.

If $\varphi\in \coinf(\rr)$ with $\varphi\geq0, \int \varphi (s)ds=1$, setting $\varphi_{n}(s)= n \varphi(ns)$, we can compute $Df$  for $f\in \coinf(S_{i})\otimes \cc^{2}$ as
\begin{equation}
\label{e2.001}
Df= \lim_{n\to +\infty}\gamma\circ G(  \varphi_{n}(s)\otimes (f_{1}(y)-r_{0}(y)f_{0}(y))- \varphi_{n}'(s)\otimes f_{1}(y)),
\end{equation}
where the limit takes place in $ \cinf(S)\oplus \cinf(S)$.
  
Note that it is not obvious that $Df\in \cinf(S)\oplus \cinf(S)$. To prove it one can first replace $G$ by a properly supported pseudodifferential parametrix $P^{(-1)}\in \Psi^{-2}_{\rm c}(N)$. Using then Gaussian normal coordinates near a point $x^{0}\in S$, one is reduced locally to $N= \rr^{d}$, $S= \{x_{1}=0\}$. The details can be found for example in \cite[Sects. 5.1- 5.3]{CP}.
 
 Another useful identity is the following:  for $u\in \cinf(\overline{\Omega})$ let  $I u$ be the extension of $u$ by $0$ in $N\backslash \overline{\Omega}$. Then 
 \begin{equation}
\label{e2.4}
PIu= \tilde{f}_{1}(dVol_{h})^{-1}dS- \tilde{f}_{0}(dVol_{h}^{-1})\p_{\nu}^{*}dS+ I Pu, \hbox{ for }f= \gamma u.
\end{equation}
 \section{The double $\beta-$KMS state}\label{sec4}\init
 In this section we consider the {\em double} $\beta-${\em KMS state}  $\omega_{\beta}$  in ${\mathcal M}^{+}\cup {\mathcal M}^{-}$.  It is obtained as the natural extension to ${\mathcal M}^{+}\cup {\mathcal M}^{-}$ of the  state $\omega_{\beta}^{+}$ in $\mathcal{M}^{+}$, which is a $\beta-$KMS  state in ${\mathcal M}^{+}$ with respect to the Killing flow . Its construction, for the more general stationary case is given in \cite[Thm. 3.5]{S}. 
 
 Since $\Sigma\backslash \cB$ is a Cauchy surface for ${\mathcal M}^{+}\cup {\mathcal M}^{-}$, we associate to $\omega_{\beta}$ its (complex) Cauchy surface covariances on $\Sigma\backslash \cB$ $\lambda^{\pm}$, and (since $\omega_{\beta}$ is a pure state), the pair of complementary projections $c^{\pm}= \pm q^{-1}\circ \lambda^{\pm}$, see Subsect. \ref{secf.1}. We will study in details the projection  $c^{+}$.
 
 We identify $\coinf(\Sigma\backslash \cB)$ with $\coinf(\Sigma^{+})\otimes \cc^{2}$ using the map
 \beq\label{defdeRhat}
\begin{array}{rl}
\hat{R}:& \coinf(\Sigma^{+})\otimes \cc^{2}\to \coinf(\Sigma^{+})\oplus \coinf(\Sigma^{-})\\[2mm]
&g= g^{(0)}\oplus g^{(\beta/2)}\mapsto f= g^{(0)}\oplus r^{*}g^{(\beta/2)},
\end{array}
\eeq
where $r: \Sigma\to \Sigma$ is the restriction to $\Sigma$ of the wedge reflection $R$, see Subsect. \ref{sec1.2}.

We will show that 
\[
C\defeq \hat{R}^{-1}\circ c^{+}\circ \hat{R}
\]
is exactly the Calder\'{o}n projector  for the Euclidean operator  $K_{+}$ acting on $(N, \hat{g})$, see Subsect. \ref{sec2.1}, and the open set 
 \[
\Omega\defeq \{(\tau, y)\in N : 0<\tau<\beta/2\}.
\]
 \subsection{The double $\beta-$KMS state}\label{sec4.1}
 We recall now the expression of $\omega_{\beta}$ given by Sanders, see \cite[Sect. 3.3]{S}.  
 
 There are some differences in signs and factors of $\i$ with the expression given by Sanders in  \cite[Sect. 3.3]{S}.  They come from two differences between our convention for quantized Klein-Gordon fields and the one of Sanders:
 
 - our convention for Cauchy data of   a solution of $Pu=0$  is given the map $\rho u= \col{u\tra{\Sigma}}{\i^{-1}\p_{\nu}u\tra{\Sigma}}=: f$,  which is more natural for complex fields and leads to a more symmetric formulation of the Hadamard condition, while Sanders uses   $\rho u= \col{u\tra{\Sigma}}{\p_{\nu}u\tra{\Sigma}}=g$, so $f= \mat{1}{0}{0}{-\i}g$.
  
  - we use as  complex symplectic form     $\bar{f}\dito \sigma f'= (f_{1}|f_{0}')- (f_{0}| f'_{1})$, while Sanders uses  $g\dito \sigma g'= (g_{0}|g_{1}')- (g_{1}| g_{0}')$. In terms of spacetime fields, we use $i^{-1}E$, Sanders uses $\i E$.

 Let us unitarily identify $L^{2}(\Sigma, |h|^{\12}dy)$ with $L^{2}(\Sigma^{+}, |h|^{\12}dy)\oplus L^{2}(\Sigma^{-}, |h|^{\12}dy)$, by
 \[
u\mapsto u_{+}\oplus u_{-}, \ u_{\pm}= u\tra{\Sigma^{\pm}}.
\]
Under this identification the action of the wedge reflection $r^{*}u= u\circ r$ will be denoted by $T$, with:
\beq\label{e1.1c}
T(u_{+}\oplus u_{-})\defeq r^{*}u_{-}\oplus r^{*}u_{+}.
\eeq
 A direct comparison with the formulas in \cite[Sect. 3.3]{S}, using the identity \eqref{ef.2} gives the following proposition.
\begin{proposition}
 The double $\beta-$KMS state on ${\mathcal M}^{+}\cup {\mathcal M}^{-}$ is given by the Cauchy surface covariance $\lambda^{+}= \mat{\lambda^{+}_{00}}{\lambda^{+}_{01}}{\lambda^{+}_{10}}{\lambda^{+}_{11}}$ where:
 \beq\label{e2.1}
\begin{array}{l}
\lambda^{+}_{00}= \12 |v|^{\12}\left(\epn^{-1}\cotn+ \epn^{-1}T \shn\right)|v|^{\12},\\[2mm]
\lambda^{+}_{11}=  \12 |v|^{-\12}\left(\epn\cotn- \epn T \shn\right)|v|^{-\12},\\[2mm]
\lambda^{+}_{01}=\lambda^{+}_{10}= \12 \one.
\end{array}
\eeq
\end{proposition}
As in Subsect. \ref{secf.1} we have $\lambda^{-}= \lambda^{+}-q$, where 
 the charge $q= \i\sigma $ is given by the matrix $q= \mat{0}{1}{1}{0}$.  We introduce the operators $c^{\pm}\defeq\pm q^{-1}\lambda^{\pm}$ and obtain
 \begin{equation}
\label{e2.2}
 c^{+}= \mat{\12}{\lambda^{+}_{00}}{\lambda^{+}_{11}}{\12}.
\end{equation}
 Note that if
\[
b_{0}=\epn^{-1}\cotn+ \epn^{-1}T \shn, \ b_{1}= \epn\cotn- \epn T \shn,
\]
then using that $[T, \epsilon]=0$ we obtain that
\[
b_{0}b_{1}= b_{1}b_{0}= \cotn^{2}- \shn^{2}= 1, 
\]
from which it follows easily that $c^{\pm}$ are (formally) projections. This is expected since  the double $\beta-$KMS state $\omega_{\beta}$ is
 a pure state in ${\mathcal M}^{+}\cup {\mathcal M}^{-}$.
 \subsection{Conjugation by $\hat{R}$}\label{sec4.2}
 The map $\hat{R}$ defined in \eqref{defdeRhat} allows to unitarily identify $L^{2}(\Sigma^{+})\otimes \cc^{2}$ with $L^{2}(\Sigma^{+})\oplus L^{2}(\Sigma^{-})$. We have:
\begin{equation}
\label{e2.4b}
\hat{R}^{-1}\epsilon  \hat{R}= \epsilon_{+}\oplus \epsilon_{+}, \ \hat{R}^{-1} T  \hat{R}= \mat{0}{1}{1}{0}.
\end{equation}
Denoting by $c^{+}_{ij}$ for $i,j\in \{0, 1\}$ the entries of  the matrix $c^{+}$ and setting 
\[
C_{ij}\defeq \hat{R}^{-1}\circ c^{+}_{ij}\circ \hat{R},
\]
\def\av{|v|}
we obtain after an easy computation using \eqref{e2.1}, \eqref{defdeRhat}:
\begin{equation}
\label{e2.5}
\begin{array}{rl}
C_{00} g_{0}=& \12 g_{0}^{(0)}\oplus \12 g_{0}^{(\beta/2)}, \\[2mm]
C_{11}g_{1}=&\12 g_{1}^{(0)}\oplus \12 g_{1}^{(\beta/2)}, \\[2mm]
C_{01}g_{1}=&\12|v|^{\12}\epp^{-1}\cotp |v|^{\12}g_{1}^{(0)}+ \12
|v|^{\12}\epp^{-1}\shp |v|^{\12}g_{1}^{(\beta/2)}\\[2mm]
&\oplus \12|v|^{\12}\epp^{-1}\cotp |v|^{\12}g_{1}^{(\beta/2)}+ \12
|v|^{\12}\epp^{-1}\shp |v|^{\12}g_{1}^{(0)},\\[2mm]
C_{10}g_{0}=& \12 |v|^{-\12} \epp\cotp |v|^{-\12}g_{0}^{(0)}- \12 |v|^{-\12}\epp \shp |v|^{-\12}g_{0}^{(\beta/2)}\\[2mm]
&\oplus \12 |v|^{-\12} \epp\cotp |v|^{-\12}g_{0}^{(\beta/2)}- \12 |v|^{-\12}\epp \shp |v|^{-\12}g_{0}^{(0)}.
\end{array}
\end{equation}
In \eqref{e2.5} the upper indices $(0)$, $(\beta/2)$ refer to the two connected components $\{\tau=0\}$ and $\{\tau= \beta/2\}$ of $\p\Omega$, while the lower indices $0$, $1$ refer to the two components of $g$. \subsection{The Calder\'{o}n projector}\label{sec4.3}
 We now compute the Calder\'{o}n projector for $K_{+}$, associated to the Riemannian manifold $(N, \hat{g})$. We choose
 \[
\Omega= \{(\tau, y)\in N : 0<\tau<\beta/2\}.
\]
\def\be2{\frac{\beta}{2}}
 We have $S= \p \Omega= S_{0}\cup S_{\beta/2}$ and we write  $f\in \coinf(S)\oplus\coinf(S)$ as $f= f^{(0)}\oplus f^{(\beta/2)}$ for $f^{(i)}\in  \coinf(S_{i})\oplus\coinf(S_{i})$.  
 
 We denote by $\gamma^{(i)}$, $i= 0, \beta/2$ the trace operators on $S_{i}$ defined by $\gamma u = \gamma^{(0)}u\oplus \gamma^{(\beta/2)}u$  for $u\in \cinf(\overline{\Omega})$. We have: 
 \begin{equation}
\label{e4.1}\begin{array}{l}
\gamma^{(0)}u= \lim_{\tau\to 0^{+}}\col{u(\tau, y)}{- |v(y)|^{-1}\p_{\tau}u(\tau, y)}, \\[2mm]
 \gamma^{(\beta/2)}u= \lim_{\tau\to (\beta/2)^{-}}\col{u(\tau, y)}{|v(y)|^{-1}\p_{\tau}u(\tau, y)}.
\end{array}
\end{equation}
 We denote similarly by $\p_{\nu}^{(i)}$ the exterior normal derivatives on $S_{i}$.

We compute the Calder\'{o}n projector $D$  defined in Subsect. \ref{sec3.2} using the coordinates $(\tau, y)$. Since  $dS_{i}= |h|^{\12}(y)dy$ and $dVol_{\hat{g}}= |v|^{\12}(y)|h|^{\12}(y)dy$, we obtain:
\begin{equation}
\label{e4.2}
Df= D^{(0)}f\oplus D^{(\beta/2)}f,
\end{equation}
for
\begin{equation}
\label{e4.3}
\begin{array}{rl}
D^{(i)}f
=&\gamma^{(i)}\circ G\circ |v|^{-1}\left(\p_{\nu}^{(0)}\delta_{0}(\tau)\otimes f^{(0)}_{0}(y)+ \delta_{0}(\tau)\otimes f^{(0)}_{1}(y)\right.\\[2mm]
&\left.+ \p_{\nu}^{(\beta/2)}\delta_{\beta/2}(\tau)\otimes f^{(\beta/2)}_{0}(y)+ \delta_{\beta/2}(\tau)\otimes f^{(\beta/2)}_{1}(y)\right).
\end{array}
\end{equation}
Since $G= |v|^{\12}\tilde{G}|v|^{3/2}$ we have $G\circ |v|^{-1}= |v|^{\12}\tilde{G}|v|^{\12}$.  Denoting by $D^{(i)(j)}_{kl}$ for $i, j\in \{0, \beta/2\}$ and $k,l\in \{0, 1\}$ the various entries of $D$, we obtain:
\begin{equation}
\label{e4.4}
D^{(i)(j)}_{kl}v=\left\{\begin{array}{l}
\gamma^{(i)}_{k}|v|^{\12}\tilde{G}|v|^{\12}(\p_{\nu}^{(j)}\delta_{j}(\tau)\otimes v(y)), \ l=0,\\
\gamma^{(i)}_{k}|v|^{\12}\tilde{G}|v|^{\12}(\delta_{j}(\tau)\otimes v(y)), \ l=1.
\end{array}\right.
\end{equation}
 We also set
 \[
\p_{\tau}^{(i)}=\mp \p_{\tau}, \hbox{ for }i= 0, \beta/2,
\]
so that $\p_{\nu}^{(i)}= |v|^{-1}(y)\p_{\tau}^{(i)}$.  
\begin{proposition}\label{propokey}
 We have $D= \hat{R}^{-1}\circ c^{+}\circ \hat{R}$.
\end{proposition}
\proof 
We recall that $C_{ij}$ are the entries of $\hat{R}^{-1}\circ c^{+}\circ \hat{R}$.
We  compute $D^{(i)(j)}_{kl}$ using \eqref{e4.4} and the explicit formulas \eqref{e1.3}, \eqref{e1.4} for the kernel $\tilde{G}(\tau, \tau')$ of $\tilde{G}$.
\def\cp{|v|^{\12}}
\def\cm{|v|^{-\12}}
\def\tG{\tilde{G}}
\def\b2{\beta/2}
{\it Computation of }$D_{00}$: 
\[
\begin{array}{rl}
&D^{(0)(0)}_{00}u= \gamma^{(0)}_{0}\cp \tG \cm \p_{\tau}^{(0)}\delta_{0}\otimes u\\[2mm]
=&  \cp \lim_{\tau\to 0^{+}}\p_{\tau'}\tG(\tau, 0)\cm u= \12 u,\\[2mm]
&D^{(0){(\b2)}}_{00}u= \gamma^{(0)}_{0}\cp \tG \cm \p_{\tau}^{(\b2)}\delta_{\b2}\otimes u\\[2mm]
=&  -\cp \lim_{\tau\to 0^{+}}\p_{\tau'}\tG(\tau, \b2)\cm u=0,\\[2mm] 
&D^{(\b2)(0)}_{00}u= \gamma^{(\b2)}_{0}\cp \tG \cm \p_{\tau}^{(0)}\delta_{0}\otimes u\\[2mm]
=&  \cp\lim_{\tau\to \b2^{-}}\p_{\tau'}\tG(\tau, 0)u=0,\\[2mm]
&D^{(\b2)(\b2)}_{00}u= \gamma^{(\b2)}_{0}\cp \tG \cm  \p_{\tau}^{(\b2)}\delta_{\b2}\otimes u\\[2mm]
=& - \cp \lim_{\tau\to \b2^{-}}\p_{\tau'}\tG(\tau, \b2)\cm u= \12 u.
\end{array}
\]
Hence
\[
D_{00}g_{0}=  C_{00}g_{0}.
\]
{\it Computation of }$D_{11}$: 
\[
\begin{array}{rl}
&D_{11}^{(0)(0)}u= \gamma_{1}^{(0)}\cp\tG \cp \delta_{0}\otimes u\\[2mm]
=&  -\cm \lim_{\tau\to 0^{+}}\p_{\tau} \tG(\tau, 0)\cp u = \12 u,\\[2mm]
&D_{11}^{(0)(\b2)}u= \gamma_{1}^{(0)}\cp \tG\cp\delta_{\b2}\otimes u\\[2mm]
=& - \cm \lim_{\tau\to 0^{+}}\p_{\tau} \tG(\tau, \b2)\cp u =0,\\[2mm]
&D_{11}^{(\b2)(0)}u= \gamma_{1}^{(\b2)}\cp \tG\cp \delta_{0}\otimes u\\[2mm]
=& \cm  \lim_{\tau\to \b2^{-}}\p_{\tau}\tG(\tau, 0)\cp u=0,\\[2mm]
&D_{11}^{(\b2)(\b2)}u=  \gamma_{1}^{(\b2)}\cp \tG\cp \delta_{\b2}\otimes u\\[2mm]
=& \cm  \lim_{\tau\to \b2^{-}}\p_{\tau}\tG(\tau, \b2)\cp u= \12 u.
\end{array}
\]
Hence
\[
D_{11}g_{0}=  C_{11}g_{0}.
\]
{\it Computation of }$D_{01}$: 
\[
\begin{array}{rl}
&D_{01}^{(0)(0)}u= \gamma_{0}^{(0)}\cp \tG\cp \delta_{0}\otimes u\\[2mm]
=&\cp  \lim_{\tau\to 0^{+}}\tG(\tau, 0)\cp u =\12\cp \epp^{-1}\cotp \cp u,\\[2mm]
&D_{01}^{(0)(\b2)}u= \gamma_{0}^{(0)}\cp \tG \cp \delta_{\b2}\otimes u\\[2mm]
=&\cp  \lim_{\tau\to 0^{+}}\tG(\tau, \b2)\cp u= \12\cp  \epp^{-1}\shp \cp u,\\[2mm]
&D_{01}^{(\b2)(0)}u= \gamma_{0}^{(\b2)}\cp \tG \cp \delta_{0}\otimes u \\[2mm]
=& \cp  \lim_{\tau\to \b2^{-}}\tG(\tau, 0)\cp u=\12 \cp \epp^{-1}\shp \cp  u,\\[2mm]
&D_{01}^{(\b2)(\b2)}u= \gamma_{0}^{(\b2)}\cp \tG \cp \delta_{\b2} \otimes u\\[2mm]
=&\cp  \lim_{\tau\to \b2}\tG(\tau, \b2)\cp u=\12 \cp\epp^{-1} \cotp \cp u.
\end{array}
\]
Hence 
\[
D_{01}g_{1}= C_{01}g_{1}.
\]
{\it Computation of }$D_{10}$: 
\[
\begin{array}{rl}
&D_{10}^{(0)(0)}u= \gamma_{1}^{(0)}\cp \tG \cm \p_{\tau}^{(0)}\delta_{0}\otimes u\\[2mm]
=&  -\cm   \lim_{\tau\to 0^{+}}\p_{\tau}\p_{\tau'}\tG(\tau, 0) \cm u= \12 \cm \epp \cotp \cm u,\\[2mm]
&D_{10}^{(0)(\b2)}u= \gamma_{1}^{(0)}\cp \tG \cm \p_{\tau}^{(\b2)}\delta_{\b2}u\\[2mm]
= &\cm \lim_{\tau\to 0^{+}}\p_{\tau}\p_{\tau'}\tG(\tau, \b2)\cm u=-\12\cm\epp \shp\cm u,\\[2mm]
&D_{10}^{(\b2)(0)}u= \gamma_{1}^{(\b2)}\cp \tG \cm\p_{\tau}^{(0)}\delta_{0}\otimes u\\[2mm]
=&  \cm  \lim_{\tau\to \b2^{-}}\p_{\tau}\p_{\tau'}\tG(\tau, 0)\cm u= - \12\cm \epp\shp \cm u,\\[2mm]
&D_{10}^{(\b2)(\b2)}u= \gamma_{1}^{(\b2)}\cp \tG\cp \p_{\nu}^{(\b2)}\delta_{\b2}\otimes u\\[2mm]
=& -\cm \lim_{\tau\to \b2^{-}}\p_{\tau}\p_{\tau'}\tG(\tau, \b2)\cm u= \12 \cm \epp \cotp \cm u.
\end{array}
\]
Hence
\[
D_{10}g_{0}= C_{10}g_{0}.
\]
This completes the proof of the proposition. \qed 
 
 \section{The Hartle-Hawking-Israel state and its properties}\label{final}\init
 
 \subsection{The smooth extension of $(N, \hat{g})$ and the Hawking temperature}
 The  existence of the Hartle-Hawking-Israel state and the definition of the Hawking temperature $T_{H}= \kappa(2\pi)^{-1}$ (where $\kappa$ is the surface gravity) rely on the existence of an  extension $(N_{\rm ext}, \hat{g}_{\rm ext})$ of $(N, \hat{g})$ such that  the two components $S_{0}, S_{\beta/2}\sim \Sigma^{+}$ of $\p \Omega$ are smoothly glued together into $\Sigma\subset N_{\rm ext}$.

 The extended Riemannian metric $\hat{g}_{\rm ext}$ is smooth iff $\beta= (2\pi)\kappa^{-1}$ (for other values of $\beta$ $(N_{\rm ext}, \hat{g}_{\rm ext})$ has a conic singularity on $\cB$).

Let us embed $\Sigma\backslash \cB$ into $N$ by:
\[
\hat{r}: \left\{\begin{array}{l}
x\mapsto (0, x)\hbox{ for }x\in \Sigma^{+},\\
x\mapsto (\beta/2, r(x))\hbox{ for }x\in \Sigma^{-},
\end{array}\right.
\]
 Note that  for $\hat{R}$  defined in  \eqref{defdeRhat} we have
 \begin{equation}\label{e5.2}
\hat{R}= (\hat{r})^{*}.
\end{equation}
We recall that the function $m: \Sigma\to \rr^{+}$ was introduced in Subsect. \ref{sec1.3}.
 \begin{proposition}\cite[Subsect. 2.2]{S}\label{sanders}
 Assume that $\beta= (2\pi)\kappa^{-1}$. Then there exists a smooth complete Riemannian manifold $(N_{\rm ext}, \hat{g}_{\rm ext})$ and
 \ben
 \item a smooth isometric embedding $\psi: \Sigma\to N_{\rm ext}$,
 \item a smooth isometric embedding $\chi: (N, \hat{g})\to (N_{\rm ext}\backslash \cB_{\rm ext}, \hat{g}_{\rm ext})$ for $\cB_{\rm ext}= \psi(\cB)$,
 \item a smooth function $m_{\rm ext}: N_{\rm ext}\to \rr$ with $m_{\rm ext}\geq m_{0}^{2}>0$
 \een
 such that\[
 \psi\tra{\Sigma\backslash \cB}= \chi\circ \hat{r}, \ \psi^{*}m_{\rm ext}= m\tra{N}. 
 \]
 \end{proposition}
 \begin{figure}[H]
\includegraphics[width=0.7\linewidth]{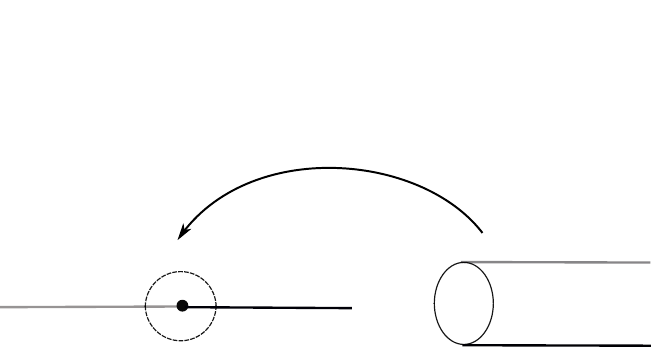}
\put(-135, 17){$\Sigma^{+}$}
\put(-220, 17){$\Sigma^{-}$}
\put(-180, 17){$\mathcal{B}$}
\put(-100, 17){$\mathbb{S}_{\beta}$}
\put(-35, 34){$\Sigma^{+}\!\!\sim \!\!r(\Sigma^{-})$}
\put(-35, 7){$\Sigma^{+}$}
\put(-67,-15){$\mathbb{S}_{\beta}\times \Sigma^{+}$}
\put(-150, -15){$\rr^{2}\times \cB$}
\put(-120, 50){$\chi$}
\caption{The embedding $\chi$}
\end{figure}
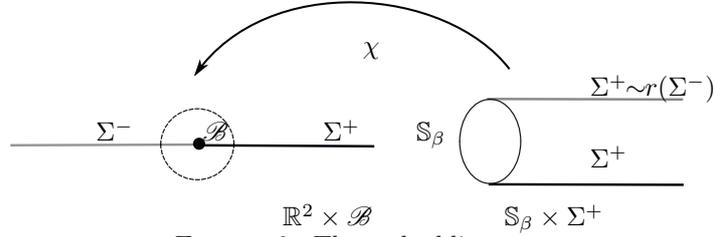

This fundamental fact is well explained in \cite[Subsect. 2.2]{S}. Let us briefly recall the construction of $\chi$ following \cite{S}:
one  introduces  Gaussian normal coordinates to $\cB$ in $(\Sigma, h)$, where $h$ is the  Riemannian metric induced by $g$ on $\Sigma$. We choose the unit normal vector field to $\cB$ pointing towards $\Sigma^{+}$. Using these coordinates we can, since $\cB$ is compact,  identify  a small neighborhood $U$ of $\cB$ in $\Sigma$  with $]-\delta, \delta[\times \cB$.  Denoting by $\omega$ local coordinates on $\cB$ we have a map
\[
\begin{array}{l}
\phi: ]-\delta, \delta[\times \cB\ni (s, \omega)\mapsto y= \exp^{h}_{\omega}(s)\in  U,\\[2mm]
U^{+}= \phi(]0, \delta[\times \cB),\\[2mm]
\phi^{*}h = ds^{2}+ k_{\alpha\beta}(s, \omega)d\omega^{\alpha}d\omega^{\beta},
\end{array}
\]
for $U^{+}=\Sigma^{+}\cap U$.
In the local coordinates $(\tau, s, \omega)$ on $\cS_{\beta}\times U^{+}$ the embedding $\chi$ takes the form:
 \beq\label{cartesian}
\chi: \begin{array}{l}
\cS_{\beta}\times ]0, \delta[\times \cB\to B_{2}(0, \delta)\times \cB\\[2mm]
(\tau, s, \omega)\mapsto (s\cos(\beta(2\pi)^{-1}\tau), s\sin(\beta (2\pi)^{-1}\tau), \omega)\eqdef(X, Y, \omega),
\end{array}
\eeq
where $B_{2}(0, \delta)= \{(X, Y)\in \rr^{2}: 0<X^{2}+ Y^{2}<\delta^{2}\}$. A straighforward computation performed in  \cite[Subsect. 2.2]{S} shows that $\hat{g}$ admits a smooth extension $\hat{g}_{\rm ext}$ to $N_{\rm ext}$ iff $\beta= (2\pi)\kappa^{-1}$.

\subsection{The extension of $\omega_{\beta}$ to $M$}
 We recall from Subsect. \ref{sec2.1} that  $K$ is defined from the  closure $\bar{Q}$ of the quadratic form $Q$  on $\coinf(N)$.   
 
 Similarly $K_{\rm ext}= - \Delta_{\hat{g}_{\rm ext}}+ m_{\rm ext}$, acting on $N_{\rm ext}$ is defined using the  corresponding quadratic form $Q_{\rm ext}$. 
 
 The following lemma is equivalent to \cite[Prop. 5.2]{S}, for completeness we give a short proof using quadratic form arguments (note that we assume the stronger condition that $\inf m(x)>0$).
 \begin{lemma}\label{l5.1}
 Let $U: \coinf(N)\to\coinf(N_{\rm ext}\backslash\cB_{\rm ext})$ defined by
 \[
Uu= u\circ \chi^{-1}.
\]
Then $U$ extends as a unitary operator $U: L^{2}(N)\to L^{2}(N_{\rm ext})$ with $K_{\rm ext}= U K U^{*}$.
\end{lemma}
 \proof $U$ clearly extends as a unitary operator. To check the second statement it suffices, taking into account the way $K$ and $K_{\rm ext}$ are defined, to prove that  $\coinf(N_{\rm ext}\backslash\cB_{\rm ext})$ is a form core for $Q_{\rm ext}$.  The domain of $\overline{Q}_{\rm ext}$ is the Sobolev space $H^{1}(N_{\rm ext})$ associated to $\hat{g}_{\rm ext}$, so we need  to show that $\coinf(N_{\rm ext}\backslash\cB_{\rm ext})$ is dense in $H^{1}(N_{\rm ext})$. Using the coordinates $(X, Y, \omega)$ near $\cB_{\rm ext}\sim \{0\}\times \cB$, this follows from the fact that $\coinf(\rr^{2}\backslash \{0\})$  is dense in $H^{1}(\rr^{2})$, see eg \cite[Thm. 3.23]{A}. \qed
 
 We recall that the projection $c^{+}$ associated to the double $\beta-$KMS state $\omega_{\beta}$ was defined in \eqref{e2.2}. Let us identify in the sequel $\Sigma$ with $\Sigma_{\rm ext}= \psi(\Sigma)\subset N_{\rm ext}$.
\begin{theorem}\label{thmain1}
 Let $D_{\rm ext}$ the Calder\'{o}n projector for $(K_{\rm ext}, \Sigma)$. Then for $f,g\in \coinf(\Sigma\backslash\cB)\otimes \cc^{2}$ we have:
 \[
(g| c^{+}g)_{L^{2}(\Sigma)}= (g_{\rm ext}| D_{\rm ext}f_{\rm ext})_{L^{2}(\Sigma)},
\]
where $f_{\rm ext}= (\psi^{*})^{-1}f$, $g_{\rm ext}= (\psi^{*})^{-1}g$.
\end{theorem}
 \proof This follows from  Prop. \ref{propokey}, the fact that $\hat{R}$ is implemented by the embedding $\hat{r}$ of $\Sigma\backslash \cB$ into $N$, (see \eqref{e5.2}) and Lemma \ref{l5.1}. \qed

\subsection{Uniqueness of the extension}

We discuss now the uniqueness of  extensions of $\omega_{\beta}$ to $M$. Other types of uniqueness results were obtained before in \cite{KW} and \cite{K}.
\begin{proposition}\label{unique}
There exists at most one quasi-free state $\omega$ for the Klein-Gordon field on $M$ such that:
\ben
\item  the restriction of $\omega$ to ${\mathcal M}^{+}\cup {\mathcal M}^{-}$ equals $\omega_{\beta}$,
\item the spacetime covariances $\Lambda^{\pm}$ of $\omega$ map $\coinf(M)$ into $\cinf(M)$.
\een
\end{proposition}
\proof
Let $\omega$ a quasi-free state for the Klein-Gordon operator $P$ in $M$, with spacetime covariances $\Lambda^{\pm}$. We assume that $\Lambda^{\pm}: \coinf(M)\to \cinf(M)$. Denoting by $\Lambda^{\pm}(x,x')$ their Schwartz kernels, we have $P(x, \p_{x})\Lambda^{\pm}(x, x')= P(x', \p_{x'})\Lambda^{\pm}(x, x')=0$, which implies that \beq\label{nico}
\WF(\Lambda^{\pm})'\subset \cN\times \cN,
\eeq 
where $\cN$ is defined in \eqref{defdechar}.
We claim that the entries $c^{\pm}_{k, k'}$, $k, k'=0,1$ of $c^{\pm}$  defined in \eqref{defdecplusmoins} map $\coinf(\Sigma)$ into $\cinf(\Sigma)$.  In fact by \eqref{nico} we have $\Lambda^{\pm}=  \Lambda^{\pm}\circ A$ modulo smoothing, where $A\in \Psi^{0}(M)$ is a pseudodifferential operator with ${\rm essupp}(A)$ included in an arbitrary small conical neighborhood of $\cN$. For $u\in \coinf(\Sigma)$ we have, modulo factors  of $\i$:
\[
c^{\pm}_{k, k'}u= \p_{t}^{k}\Lambda^{\pm}\circ A (-\p_{t}^{k'}\delta_{0}\otimes u)\tra{t=0},
\] 
see \eqref{defdecplusmoins}.
Since $\WF((-\p_{t}^{k'}\delta_{0})\otimes u) \subset N^{*}\Sigma$, where $N^{*}\Sigma\subset T^{*}M$ is the conormal bundle to $\Sigma$ and $\Sigma$ is spacelike, we have $N^{*}\Sigma\cap\cN= \emptyset$, hence $A (-\p_{t}^{k'}\delta_{0}\otimes u)\in \cinf(M)$, which proves our claim.

Let now $\omega_{i}$, $i=1,2$ be two quasi-free states as in the proposition.  Since $(u| (\Lambda_{1}^{+}- \Lambda_{2}^{+})v)_{L^{2}(M)}=0$ for $u, v\in \coinf({\mathcal M}^{+}\cup {\mathcal M}^{-})$ we obtain that
$(f| (\lambda_{1}^{+}- \lambda_{2}^{+})g)_{L^{2}(\Sigma)\otimes \cc^{2}}=0$ for $f, g\in \coinf(\Sigma\backslash \cB)\otimes \cc^{2}$ hence $\supp (\lambda_{1}^{+}- \lambda_{2}^{+})g\subset \cB$ for $g\in \coinf(\Sigma\backslash \cB)\otimes \cc^{2}$. Since we have seen that $\lambda_{i}^{+}: \coinf(\Sigma)\otimes \cc^{2}\to \cinf(\Sigma)\otimes \cc^{2}$ this implies that $(\lambda_{1}^{+}- \lambda_{2}^{+})f=0$ for $f\in \coinf(\Sigma\backslash \cB)\otimes \cc^{2}$. Since $\lambda_{i}^{+}$ are selfadjoint for $L^{2}(\Sigma, dVol_{h})\otimes \cc^{2}$ this implies that $\supp (\lambda_{1}^{+}- \lambda_{2}^{+})f\subset \cB$ for $f\in \coinf(\Sigma)\otimes \cc^{2}$, hence $(\lambda_{1}^{+}- \lambda_{2}^{+})f=0$ using again that $\lambda_{i}^{+}: \coinf(\Sigma)\otimes \cc^{2}\to \cinf(\Sigma)\otimes \cc^{2}$. \qed

\subsection{The Hartle-Hawking-Israel state}

\begin{theorem}[\cite{S}]\label{thmain2}
Let us set 
\[
\lambda^{+}_{\rm HHI}\defeq q\circ D_{\rm ext}, \ \lambda^{-}_{\rm HHI}\defeq  \lambda^{+}_{\rm HHI}-q,
\]
 where $D_{\rm ext}$ is the Calder\'{o}n projector for $(K_{\rm ext}, \Sigma)$ and the charge quadratic form $q$ is defined in \eqref{defdeq}. Then:
 \ben
\item  $\lambda^{\pm}_{\rm HHI}$ are the Cauchy surface covariances for the Cauchy surface $\Sigma$ of a quasi-free state $\omega_{\rm HHI}$ for the free Klein-Gordon field on $M$.

\item the Hartle-Hawking-Israel state $\omega_{\rm HHI}$ is a pure Hadamard state and is the unique extension to $M$ of  the double $\beta-$KMS state $\omega_{\beta}$ with the property that its spacetime covariances $\Lambda^{\pm}_{\rm HHI}$ map  continuously $\coinf(M)$ into $\cinf(M)$.
\een
\end{theorem}
\proof
Let us first prove (1).  By \eqref{ef.2bb} it suffices to check  the positivity of $\lambda^{+}_{\rm HHI}$. This was shown in \cite[Thm. 5.3]{S} using reflection positivity. For the reader's convenience, let us briefly repeat the argument:

for $u\in L^{2}(N)$ we set $Ru(\tau, y)= u(-\tau, y)$, for $\tau\in [-\beta/2, \beta/2]\sim \bS_{\beta}$. The operator $G= K^{-1}$ is {\em reflection positive}, i.e.
\begin{equation}
\label{rp}
(Ru| Gu)_{L^{2}(N)}\geq 0, \ \forall u\in L^{2}(N), \ \supp u\subset [0, \beta/2]\times \Sigma^{+}.
\end{equation}
 In fact setting $\tilde{u}= |v|^{3/2}u$, \eqref{rp} is equivalent to
 \begin{equation}
 \label{rpt}
 (R\tilde{u}| \tilde{G}\tilde{u})_{L^{2}(\bS_{\beta})\otimes L^{2}(\Sigma^{+})}\geq 0,  \ \forall \tilde{u}\in L^{2}(\bS_{\beta})\otimes L^{2}(\Sigma^{+}), \ supp \tilde{u}\subset [0, \beta/2]\times \Sigma^{+}.
 \end{equation}
 Using \eqref{e1.3} we obtain
 \[
 \begin{array}{rl}
 &(R\tilde{u}| \tilde{G}\tilde{u})_{L^{2}(\bS_{\beta})\otimes L^{2}(\Sigma^{+})}\\[2mm]
 =&(u_{0}| \frac{1}{2\epsilon(1- \e^{- \beta \epsilon})}u_{0})_{L^{2}(\Sigma^{+})}+ (u_{\beta}| \frac{1}{2\epsilon(1- \e^{- \beta \epsilon})}u_{\beta})_{L^{2}(\Sigma^{+})},
 \end{array}
 \]
 for
 \[
 u_{0}= \int_{S_{\beta}} \e^{- \tau \epsilon}\tilde{u}(\tau)d\tau, \ u_{\beta}= \int_{S_{\beta}} \e^{(\tau- \beta/2) \epsilon}\tilde{u}(\tau)d\tau
 \]
 where $\tilde{u}$ is identified with  the map $\bS_{\beta}\ni \tau\mapsto \tilde{u}(\tau)\in L^{2}(\bS_{\beta}; L^{2}(\Sigma^{+}))$. This proves \eqref{rp}.  
 
 By Lemma \ref{l5.1} and using that $G_{\rm ext}= K_{\rm ext}^{-1}$ is bounded on $L^{2}(N_{\rm ext})$, we deduce from \eqref{rp} that $G_{\rm ext}$ is also reflection positive, ie
 \begin{equation}
 \label{rpext}
 (R_{\rm ext}u| G_{\rm ext}u_{\rm ext})_{L^{2}(N_{\rm ext})}\geq 0, \  u\in L^{2}(N_{\rm ext}), \ \supp u\subset N_{\rm ext}^{+}= \chi([0, \beta/2]\times \Sigma^{+}),
 \end{equation}
 for $R_{\rm ext}= U RU^{*}$. By the remark before \cite[Thm. 5.3]{S}, if $(s, y)$ are Gaussian normal coordinates to $\Sigma$ in $N_{\rm ext}$ we have $R_{\rm ext}u(s, y)= u(-s, y)$, ie $R_{\rm ext}$ is given by the reflection in Gaussian normal coordinates.  This map is a isometry of $(N_{\rm ext}, \hat{g}_{\rm ext})$, which implies that if $\hat{g}_{\rm ext}= ds^{2}+ h_{\rm ext}(s, y)dy^{2}$ near $\Sigma$, we have $h_{\rm ext}(s, y)= h_{\rm ext}(-s, y)$ hence if $r_{s}(y)= |h_{\rm ext}(s, y)|^{-\12}\p_{s} |h_{\rm ext}(s, y)|^{\12}$ we have $r_{0}(y)\equiv 0$. 
 
 If $f\in \coinf(\Sigma)\otimes \cc^{2}$ it follows from \eqref{defdet} that $D_{\rm ext}f= \gamma G_{\rm ext} Tf$ for
 \[
 Tf= \delta_{0}(s)\otimes f_{1}- \delta_{0}'(s)\otimes f_{0}.
 \]
 We have $R_{\rm ext}Tf= \delta_{0}(s)\otimes f_{1}+ \delta_{0}'(s)\otimes f_{0}$.
 Applying the reflection positivity \eqref{rpext} to $u= Tf$ we obtain that:
 \[
 (R_{\rm ext}Tf| G_{\rm ext}Tf)_{L^{2}(N_{\rm ext})}= (f| q D_{\rm ext}f)_{L^{2}(\Sigma)}\geq 0,
 \]
 which proves the positivity of $\lambda^{+}_{\rm HHI}$. To make the argument  rigorous is suffices to approximate $\delta_{0}$ by a sequence $\varphi_{n}$ as in \eqref{e2.001}.  This completes the proof of (1).

Let us now prove (2). The fact that $\omega_{\rm HHI}$ is the unique extension of $\omega_{\beta}$ to $M$ with the stated properties has been proved in Prop. \ref{unique}. It remains to prove that $\omega_{\rm HHI}$ is a pure Hadamard state in $M$.

The fact that $\omega_{\rm HHI}$ is pure follows from the fact that $D_{\rm ext}$ is a projection. 
To prove the Hadamard property let us fix a reference Hadamard state $\omega_{\rm ref}$ for the Klein-Gordon field in $M$. By Thm. \ref{allhad} its Cauchy surface covariances on $\Sigma$ $\lambda^{\pm}_{\rm ref}$ are matrices of pseudodifferential operators on $\Sigma$. The same is true of $c^{\pm}_{\rm ref}=\pm q^{-1}\circ \lambda^{+}_{\rm ref}$ and of $c^{\pm}_{\rm HHI}$, since Calder\'{o}n projectors are given by matrices of pseudodifferential operators on $\Sigma$.

Moreover we know  that the restriction of $\omega_{\rm HHI}$ to ${\mathcal M}^{+}\cup {\mathcal M}^{-}$ is a Hadamard state. The same is obviously true of the restriction of $\omega_{\rm ref}$ to ${\mathcal M}^{+}\cup {\mathcal M}^{-}$. Going to Cauchy surface covariances, this implies that if $\chi\in \coinf(\Sigma^{\pm})$ then
\[
\chi\circ (c^{\pm}_{\rm HHI}- c^{\pm}_{\rm ref})\circ \chi\hbox{ is a smoothing operator on }\Sigma.
\]
We claim that this implies that $c^{\pm}_{\rm HHI}- c^{\pm}_{\rm ref}$ is smoothing, which will imply that $\omega_{\rm HHI}$ is a Hadamard state. 

If fact let  $a$ be one of the entries of $c^{\pm}_{\rm HHI}- c^{\pm}_{\rm ref}$, which is a scalar pseudodifferential operator belonging to $\Psi^{m}(\Sigma)$ for some $m\in \rr$. We know that  $\chi\circ a\circ \chi$ is smoothing for any $\chi\in \coinf(\Sigma\backslash \cB)$.  Then its principal symbol $\sigma_{\rm pr}(a)$ vanishes on $T^{*}(\Sigma\backslash \cB)$ hence on $T^{*}\Sigma$ by continuity, so $a\in\Psi^{m-1}(\Sigma)$. Iterating this argument we obtain that $a$ is smoothing, which proves our claim and completes the proof of the theorem.  \qed

\subsection*{Acknowledgments}{It is a pleasure to thank Micha{\l} Wrochna, Ko Sanders and Francis Nier for useful discussions. We are also grateful to Ko Sanders for his kind comments on a previous version of the manuscript.}

\end{document}